\documentclass[12pt,preprint]{aastex}
\usepackage{pslatex}
\usepackage{amssymb}

\shorttitle{Particles Acceleration in Converged Shocks} \shortauthors{Wang X. et al.}

\newcommand{\raa}{{\it Res. Astron. Astrophys.}}

\def\prl{ Phys. Rev. Lett.}
\def\prd{{ Phys. Rev.} D}
\def\jgr{ J. Geophys. Res.}
\def\grl{ Geophys. Res. Lett.}
\def\ssr{ Space Sci. Rev.}
\def\apj{ ApJ}
\def\apjl{ ApJL}
\def\apjs{ ApJS}
\def\aph{ Astroparticles Physics}
\def\mnras{ MNRAS}

\def\aap{ Astron. Astrophys.}
\def\nat{ Nature}

\begin{document}

\title{Particles Acceleration in Converged Two Shocks}
\footnotetext{$*$ This work is supported by the Xinjiang Natural Science Foundation No. 2014211A069}
\author{Xin Wang\altaffilmark{1,2,3,4,5}
Joe Giacalone\altaffilmark{2} Yihua Yan\altaffilmark{3} Mingde Ding\altaffilmark{4} Na Wang\altaffilmark{1,5} Hao Shan
\altaffilmark{1,5} } \affil{\textit{1,Xinjiang Astronomical Observatory, Chinese Academy of Sciences, Urumqi 830011,
China}}
\affil{\textit{2,Lunar and Planetary Laboratory, University of Arizona, Tucson AZ 85721, American}
\affil{\textit{3,Key Laboratory of Solar Activities, National Astronomical Observatories, Chinese Academy of Sciences,
Beijing 100012, China}} \affil{\textit{4,Key Laboratory of Modern Astronomy and Astrophysics (Nanjing
University)£¬Ministry of Education, Nanjing 210093, China}} \affil{\textit{5,Key Laboratory of Radio Astronomy, Chinese
Academy of Sciences, Nanjing 210008, China}}
} \email{e-mail: wangxin@xao.ac.cn}

\begin{abstract}
{Observations show that there is a proton spectral ``break" with E$_{break}$ at 1-10MeV in some large CME-driven
shocks. Theoretical  model usually attribute this phenomenon to a diffusive shock acceleration. However, the underlying
physics of the shock acceleration still remains uncertain.  Although previous numerical models can hardly predict this
``break" due to either high computational expense or shortcomings of current models, the present paper focuses on
simulating this energy spectrum in converged two shocks by Monte Carlo numerical method. Considering the Dec 13 2006
CME-driven shock interaction with an Earth bow shock, we examine whether the energy spectral ``break" could occur on an
interaction between two shocks. As result, we indeed obtain the maximum proton energy up to 10MeV, which is the premise
to investigate the existence of the energy spectral ``break". Unexpectedly, we further find a proton spectral ``break"
appears distinctly at the energy $\sim$5MeV.}
\end{abstract}


\keywords{acceleration of particles--methods:numerical--shock waves--solar wind--Sun:coronal mass ejections(CMEs)}

\section{Introduction}
\label{sect:intro}

Diffusive shock acceleration (DSA) is regarded as the most efficient mechanism for galactic cosmic rays (CR). In the
recent years several in situ observations indicate the supernova remnants (SNRs) are the essential candidates for the
sources of the CR's proton spectrum up to the ``knee" at a few PeV.  Since the TeV $\gamma$-ray from Crab Nebula were
first clearly detected by imaging air Cherenkov telescope (IACT) in 1989, IACTs have been extensively constructed and
are operating around the world. There are about one hundred $\gamma$-ray sources are identified up to now
\citep{Amenomori09}. Among of those sources, IC443 and W44 are the highest-significance SNRs in the second Fermi-LAT
suited for a detailed study of their $\gamma$-ray spectra. IC443 and W44 are located at distances of 1.5kpc and 2.9kpc,
respectively.  Their GeV $\gamma$-ray spectra with high-energy breaks at 20GeV and 2GeV for IC443 and W44 can trace the
parent protons escaped from SNRs shock and collided with the nearby molecular clouds (MCs). In addition, both the
$\gamma$-ray emissions with a low-energy break at $\sim$ 200MeV also can be interpreted by effect of pion bump between
SNRs protons and the MCs (i.e. p+p$\rightarrow$He+$\pi^{0}$, $\pi^{0}$$\rightarrow$$2\gamma$ ) \citep{ackermann13}.
This $\gamma$-ray luminosity model can be used to explain the CRs energy spectral ``break".

In the interplanetary (IP) space, there are also find the similar energy spectral ``break" in the IP shocks. There are
six events with hard energy spectra occurred on Nov. 6, 97, Feb. 15, 01, Jan. 20, 05, Sep. 7, 05, Dec. 5, 06, and Dec.
13, 06. These six large events all have spectral ``breaks" at the energy range of $\sim$1-10MeV \citep{mewaldt08}. In
addition, there are the six largest events on the solar cycle 23 list as Jul. 14, 00, Nov. 8, 00, Sep. 24, 01, Nov. 04,
01, Nov. 22, 01, Oct. 28, 03. These events all have spectra that roll over in similar fashion beyond $\sim$50MeV. In
present paper, we discuss the Dec. 13, 06 event, which proton fluxes are based on spacecraft of ACE, STEREO, and
SAMPEX.

Although a number of in situ observations exhibit the CR's proton spectral ``break" associated with either galaxy
source or solar source, there is still no reliable prediction of this ``break" by numerical methods. According to the
theoretic model of DSA, CR's power-law spectrum span a very large energy range from 1KeV to a few 100EeV($\sim
10^{20}$eV). If one plans to simulate the total CR's energy spectrum, it's very hard for performing so expensive
computer program. In addition, numerical simulation is usual to built a simple DSA model with a short size of the
diffusive region ahead of shock. If the energy spectral ``break" associated with a large diffusive size, the simple
numerical models would hardly include this energy spectral ``break" in their simulation results.  Monte Carlo (MC)
method can easily treat thermal ion injection \citep{wy11,wwy13}. In MC method, the scattering mean free path is
assumed to be some function of the particle rigidity, so this treatment is able to follow the evolution of individual
ions long enough to model acceleration to high energies. However, the acceleration efficiency, as well as the maximum
particle energy, are limited by the finite size of the accelerating region as parameterized by the free escape boundary
(FEB). \citet{vbe08} describes the escape of particles from the SNR shock with an assumed FEB far upstream of the
shock. The distance to the FEB is a free parameter of the simulation that controls the maximum energy of accelerated
particles and the escaping CR flux. \citet{emp90} presented an ion spectra with a maximum particle energy no more than
1MeV by applying an fixed FEB ahead of the bow shock. \citet{kje96} and \citet{wy12} forward the maximum particle
energy achieving to $\sim$4MeV by applying a moving FEB with the shock. \citet{wy15} investigated that the maximum
particle energy limited in FEB can achieve a saturation at $\sim$5.5MeV using different scattering mean free path
functions.

Since the cosmic rays are important both dynamically  and diagnostically it is essential that we understand their
acceleration, transport, radiative emissions, and interaction with other components of astrophysical environments. In
particular, the CRs spectrum shape is usually referred to as a knee-ankle structure with the ``knee" at a few PeV and
the ``ankle" at a few EeV. Similarly, the IP shock energy spectrum with a ``break" at a few MeV. There are some debated
understandings for the energy spectral ``break". One of these understandings has ever proposed that the ``break"
determined by the Larmor radius for ions. A heavy nucleus with charge Z has maximum energies Z times higher than a
proton with same Larmor radius. However, the CRs consist mainly of protons with smaller numbers of other nuclei
\citep{bell13}. Another point argues that this ``break" can probably be associated with the leakage mechanism . The CRs
drive Alfven waves efficiently enough to build a transport barrier that strongly reduces the leakage leading a spectral
``break" \citep{md13}. Although the turbulent field exist a limited boundary, it would not be the main reason for
producing the energy spectral ``break". According to the DSA, the turbulent field in the precursor is generated by the
gradient of CRs pressure, which is faded gradually but not steeply. However, if a SNR shock approaches a MCs or a
pre-supernova swept-up shell with a significant amount of neutrals, confinement of accelerated particles deteriorates.
This would lead to a energy spectral ``break" \citep[see review in][]{bmr13}. In addition, there is a multiple shocks
model proposed by \citet{schneider93,mp93}. It is assumed that the medium is highly turbulent and that the number of
shocks are propagating through it. CRs particles are injected into the system and accelerated by one or more shocks
before they escaped from the system. This model may also used to study the particle spectral features such as
``breaks". In present paper, we propose that a new collided shocks model could probably inform the energy spectral
``break" in the interacted precursor regions.

How to directly follow the higher energy spectral ``tail" for understanding the ``break" is posing a challenge to the
numerical methods. Present work treats the interaction of CME-drive shock with the Earth bow shock to try investigate
the energy spectral ``break" as described in the observed Dec 13 2006 event. In converged two shocks, we apply the
Monte Carlo simulation method without FEB. There are two reasons  make it is possible to verify the spectral ``break".
Firstly, the double shocks can provide enough energy injection to produce the required highest energy ``tail".
Secondly, due to no FEB, the accelerated particles can freely interact between double shocks and renew the standard DSA
power-law shape.

In present paper, we do simulations to further investigate the energy spectrum in converged two shocks by using Monte
Carlo method. In section 2, we introduce the converged shocks model specifically. In section 3, the simulated results
and analysis are presented. In the last section, we give summaries and conclusions.

\section{Model}
\label{Model}

Halo CMEs were observed by the SOHO/LASCO coronagraphs in association with the events of 13 December 2006, with speeds
of 1774kms$^{-1}$. The flux spectra of protons in the SEP event by ACE, STEREO, and SAMPEX instruments show a ``break"
at $\sim$1-10MeV. Spectra from GOES-11 also agree over the region from 5MeV to 100MeV. Although the broken spectra
would be little debated due to system errors from multiple spacecraft, it is hard to obtain a completely spectra for a
large energy span from an individual spacecraft. So we hopefully do a simulation to obtain an entire spectra, which can
cover the energy range from  thermal energy  below $\sim$0.1MeV to superthermal energy beyond $\sim$10MeV.

\begin{figure}[t]
\centerline{
   \includegraphics[width=4in,angle=-90]{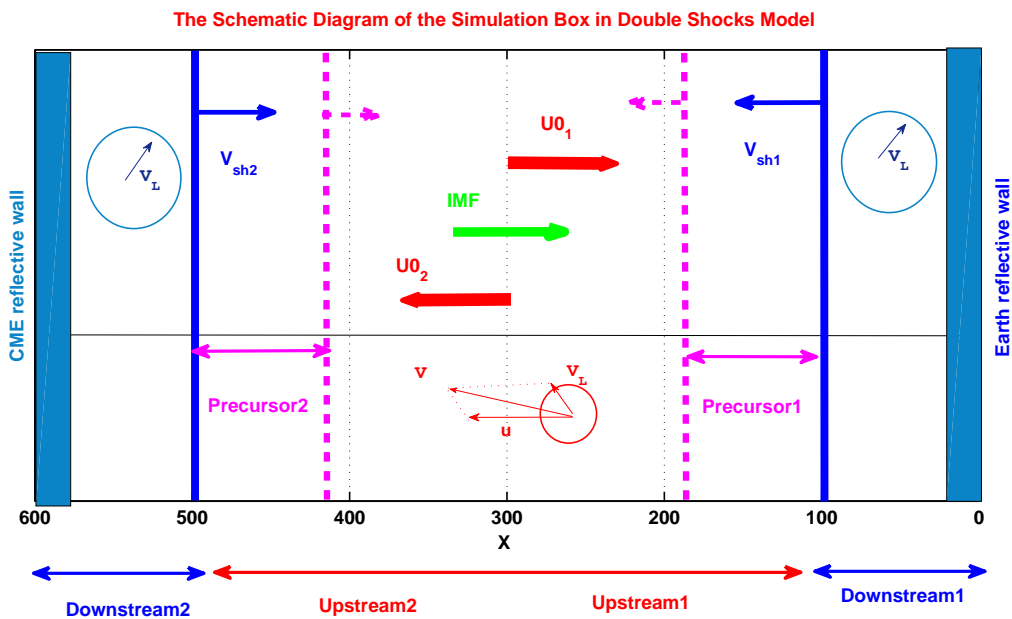}}
    \caption{\small A schematic diagram
of the simulation box. The left reflective wall represents a CME and produces one shock propagating from the left
boundary to the middle of the simulation box. The right reflective wall represents the Earth and produces another shock
involuting from the right boundary to the middle of the simulation box. Particles diffuse in the precursor regions
between two shock fronts when they approach together near the middle of the simulation box.}
  \label{fig1}
 \end{figure}

Fig.~\ref{fig1} shows a schematic diagram of the converged shocks model. The left reflective wall represents a CME,
which produces a shock by a bulk flow speed of $u0_{2}$. The right reflective wall represents the Earth, which informs
a bow shock by an opposite bulk flow speed of $u0_{1}$. Their relative speed between two bulk flows is equal to
$u=|u0_{1}|+|u0_{2}|$, which can equivalently be taken as the relative movement between two reflective walls with
opposite velocities $u0_{1}$, $u0_{2}$ in the laboratory reference frame. Similarly, we can take both downstream bulk
flow speeds with the same velocity zero in the laboratory reference frame. This model describes the double shocks
interaction occurred on the 13 December 2006 nearby Earth. According to the Wind magnetic cloud list, the cloud axis
direction is $\theta$=27$^{\circ}$ ,$\phi$=85$^{\circ}$ in GSE coordinates \citep{liu08}. In this model, we define the
bulk flow direction to the interplanetary magnetic field (IMF) direction with an oblique factor of cos($\theta$). So we
take the relative speed between two bulk flows as value for $\sim$1600km$^{-1}$ aligned to the IMF. Both two shocks are
produced by the same bulk flow speed value for $|u0_{1}|=|u0_{2}|$=$\sim$800km$^{-1}$, but with opposite direction in
the laboratory frame. Also both two reflective walls produce shocks propagating with opposite velocities of $v_{sh1}$
and $v_{sh2}$, respectively.

In this Monte Carlo method, we apply an initial number density of particles $n_{0}$ obeying a Maxwellian distribution
with a random thermal velocity $v_{0}$ in the unshocked upstream region. Initial particles with their upstream bulk
flow speeds move to their corresponding reflective walls in the both sides of the simulation box. Each bulk flow is
reflected and forms the higher shocked densities as downstream region in each side of the simulation box. With the
downstream density achieving to a saturation, the shock fronts smoothly evolute forward to the middle of the box with
shock velocity $V_{sh1}$ and $V_{sh2}$, respectively. When the two shock fronts propagate more and more close, then the
two precursors have an interaction with the time at the middle of the simulation box. It is this processes the
particles are able to gain their additional energies by crossing two shock fronts. Unlike previous single shock model,
FEB is not included in the front of the converged two shocks. Due to the two reflective walls in this model can play a
role on preventing the particles  from escaping, this can ensure the particles in upstream region have enough
opportunities of scattering on double shocks to obtain more energy gains. In this way, we can obtain the maximum
particle energy beyond 10MeV to investigate the energy spectral ``break" between 1-10MeV.

In this simulation box, the continuous bulk flow from the middle of the box with opposite directions enter into the box
along to the two boundaries of the simulation box. One bulk flow forms a CME shock in the left boundary, another bulk
flow forms a bow shock on the right boundary. With the shock formations, the particles are accelerated from each shock
front. With the shock propagations forwarding to the middle of the box, the precursor regions are mixed. This
interaction of the precursor regions ahead of two shocks make the energetic particles either re-accelerate or
de-accelerate at the end of the simulation. It is this double effect of the energetic particles is responsible to the
formation of the energy spectral ``break" at the range of 1MeV-10MeV.

In technically, the hybrid simulation method solves the equation explicitly for particle motions in an electromagnetic
plasma \citep{gbse93,gg13}. But the Monte Carlo method applies the scattering law for the particles diffusive
processes. According to the particles mean free path (mfp) equal to the local velocity times the scattering time (i.e.
$\lambda=v_{l}\cdot\tau$), we determine the scattering probabilities of particles based on the rate of the time step
over the scattering time (i.e. $\eta=dt/\tau$). So we can chose the numbers of the particles at a certain density in
each grid to diffuse with their random pitch angle deflections. By means of these diffusive processes, the particles in
upstream region transfer their kinetic energy into their random thermal energy in the downstream region. Then the
minority of these random thermal particles can diffuse back shock front by multiple scattering cycles to obtain more
energy gains to produce the superthermal energy ``tail".

The simulation parameters mainly include upstream bulk speeds of $u0_{1}$=-0.6, and $u0_{2}$=0.6, the total size of the
box $X_{max}$=600, total simulation time $T_{max}$=2400, the number of grids $n_{x}$=1200, the initial density per grid
$n_{0}$=360, the constant of the scattering time $\tau_{0}$=25/30, initial thermal velocity $\upsilon_{0}=0.02$, and
time step $dt$=1/15. The above dimensionless values are all scaled by a group of the standard scaled factors:
$x_{scale}$=2000$R_{e}$/600, $t_{scale}$=630$'$/2400, and $u_{scale}$=800kms$^{-1}$/0.6, where the $R_{e}$ is the Earth
radius. The total number of the particles in simulation box amounts to more than 1,000,000 particles.

\section{Results}
\subsection{Shock Evolution}
 \begin{figure}[t]
    \includegraphics[width=2.0in, angle=0]{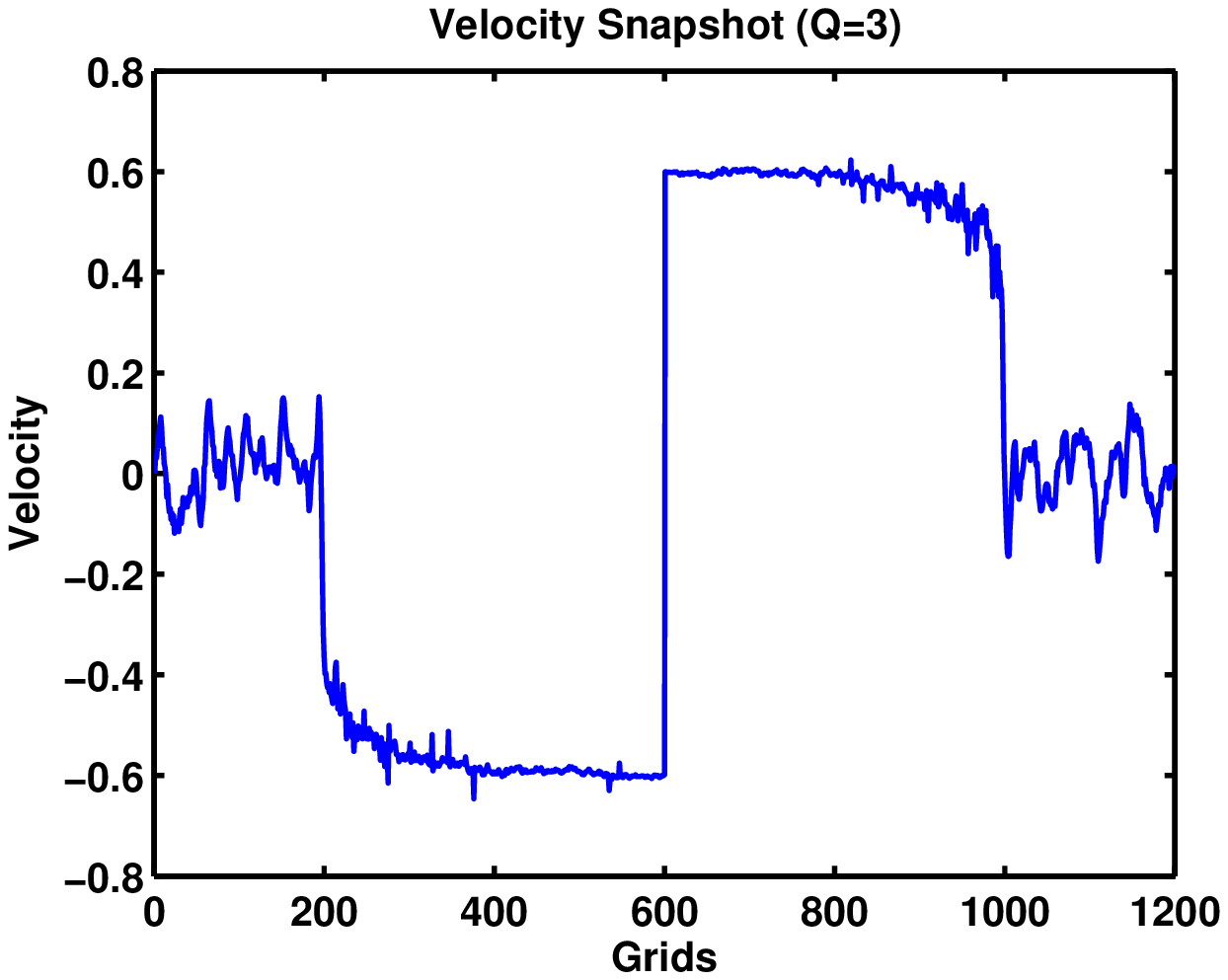}
    \includegraphics[width=2.0in, angle=0]{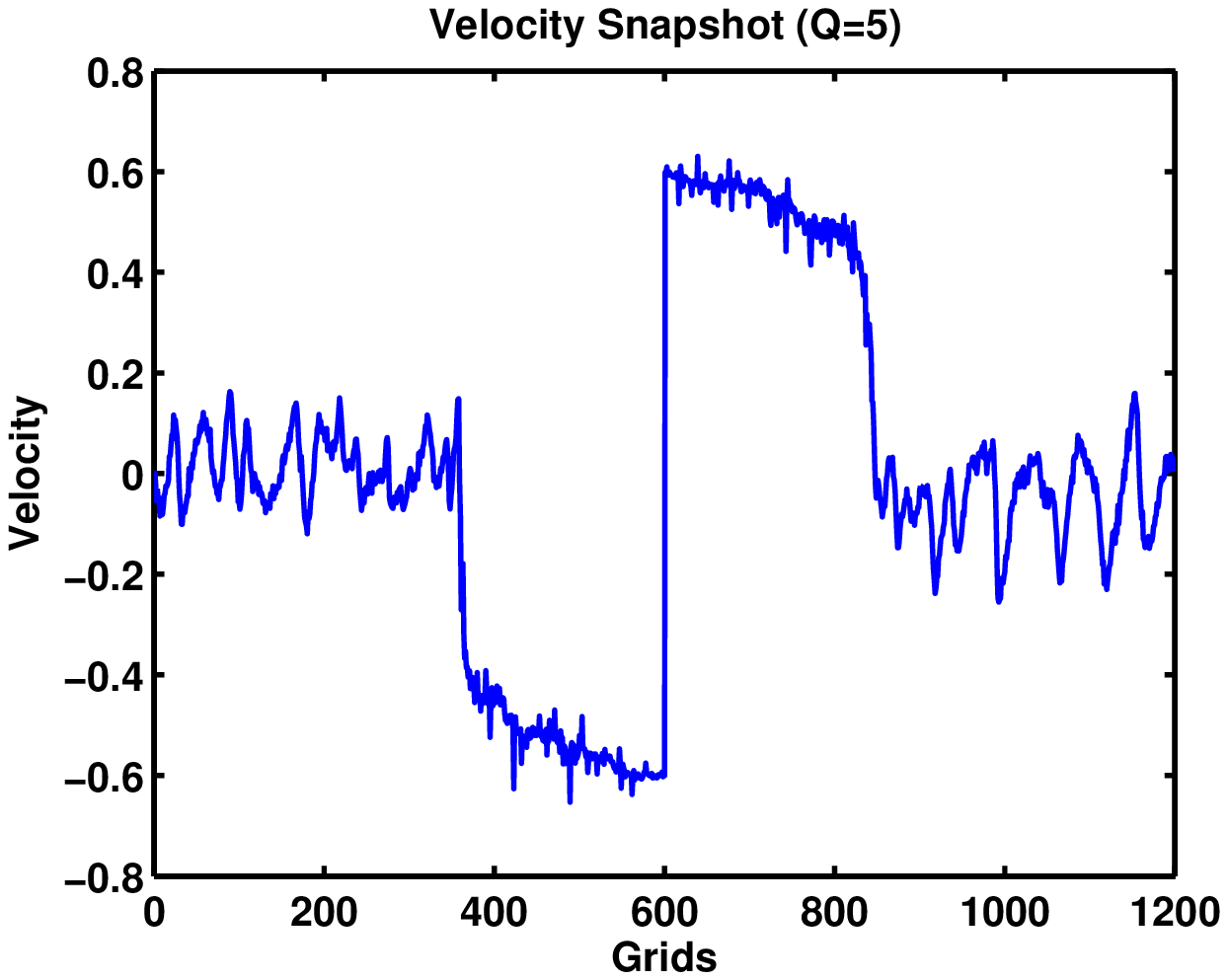}
    \includegraphics[width=2.0in, angle=0]{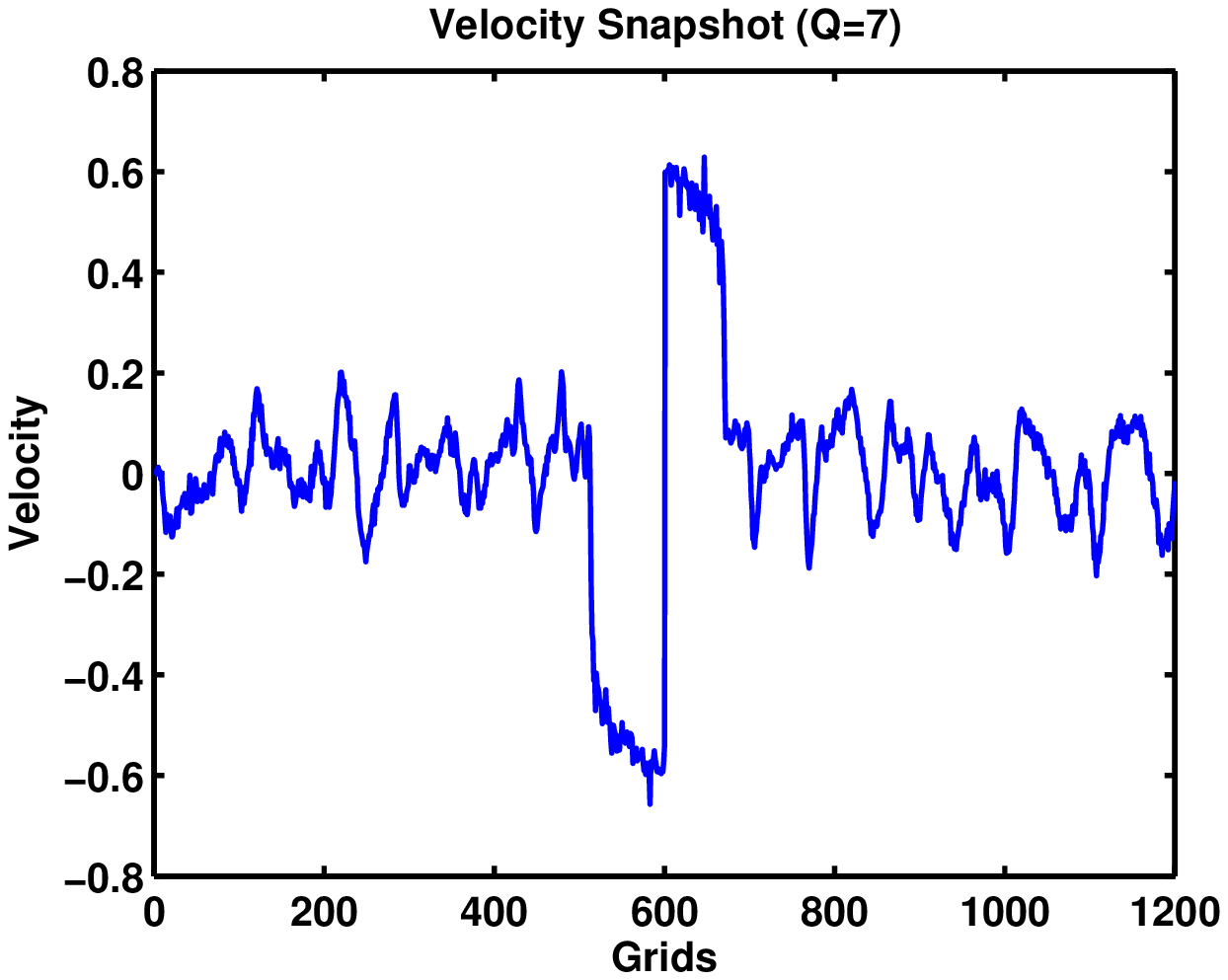}\\
    \includegraphics[width=2.0in, angle=0]{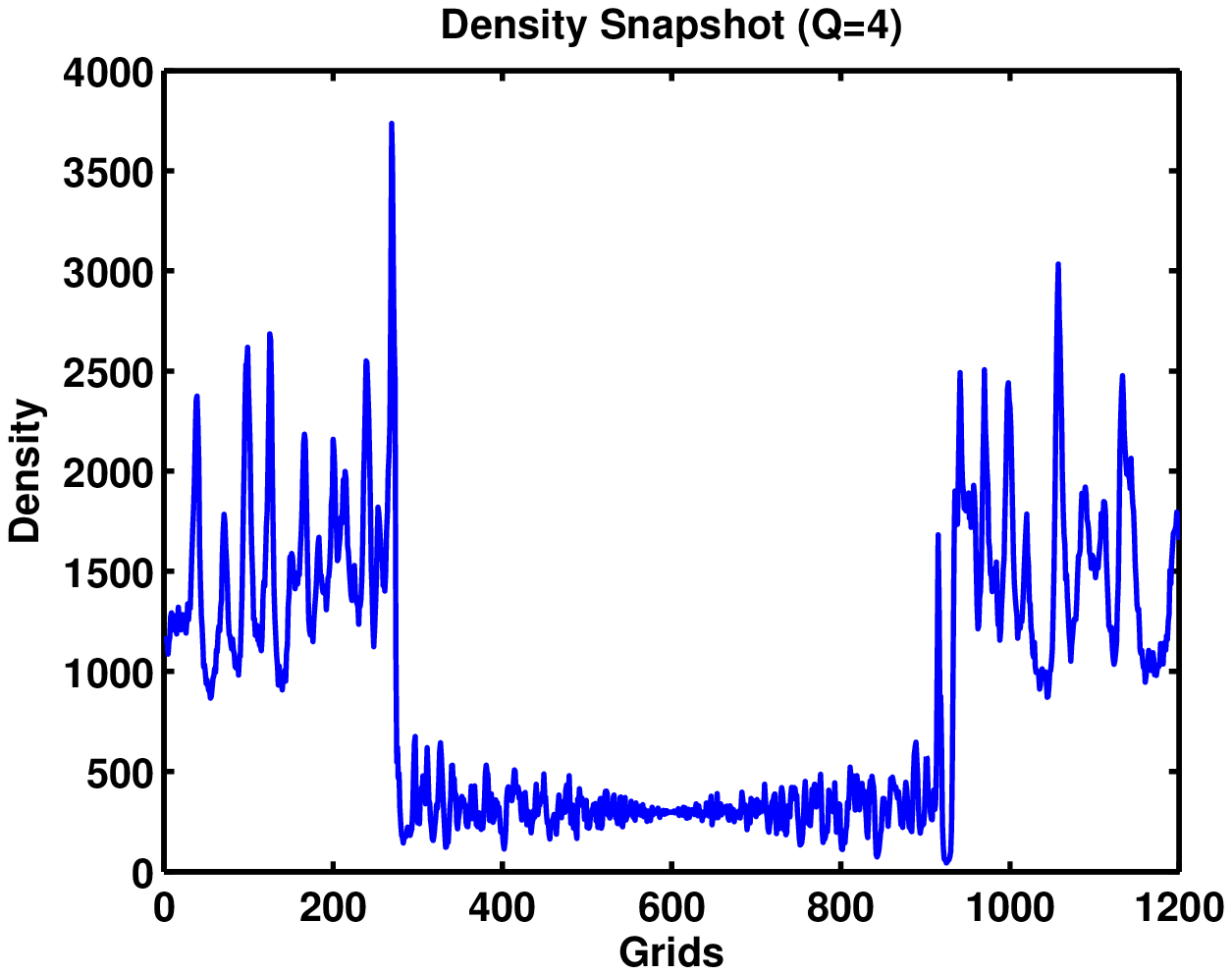}
    \includegraphics[width=2.0in, angle=0]{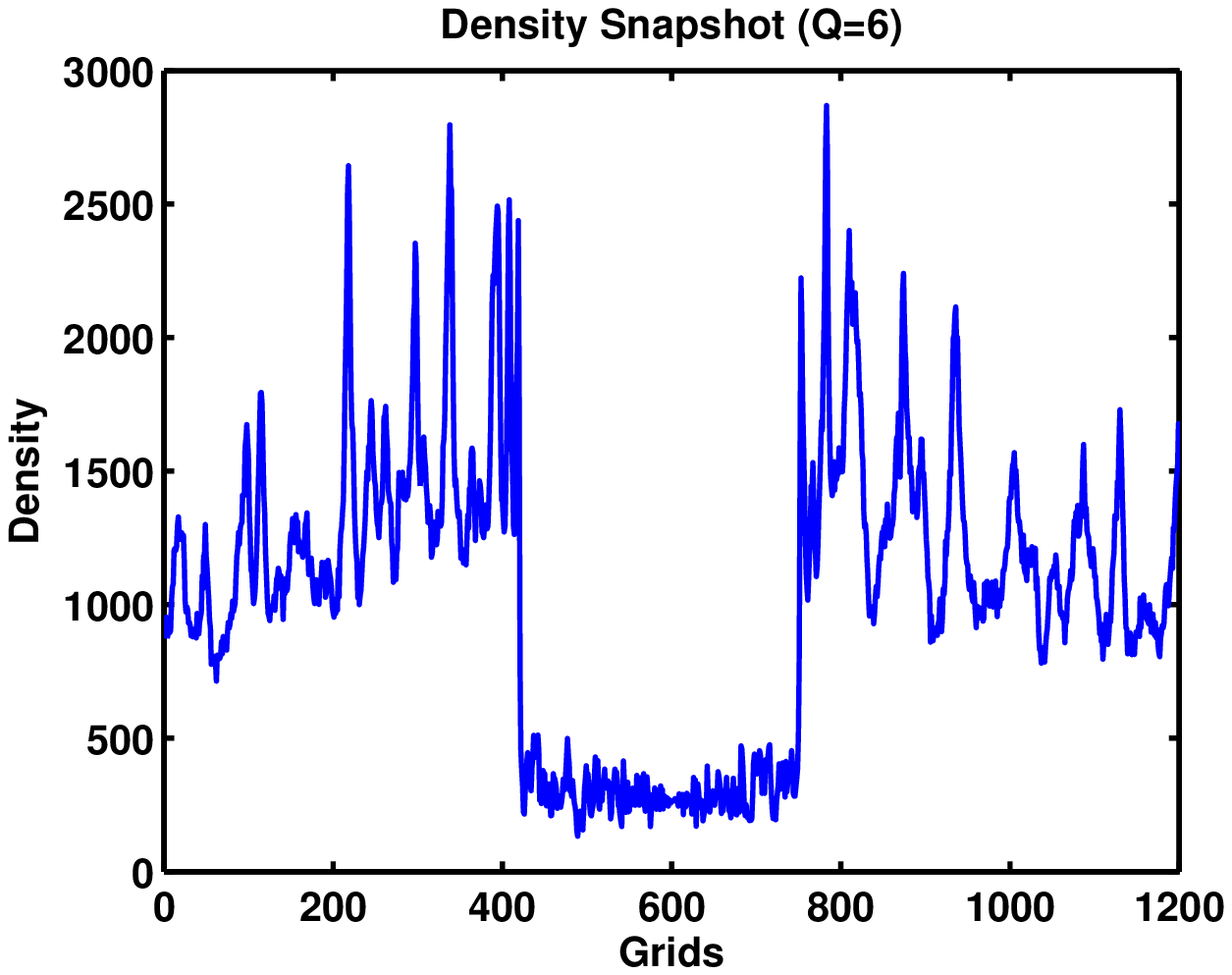}
    \includegraphics[width=2.0in, angle=0]{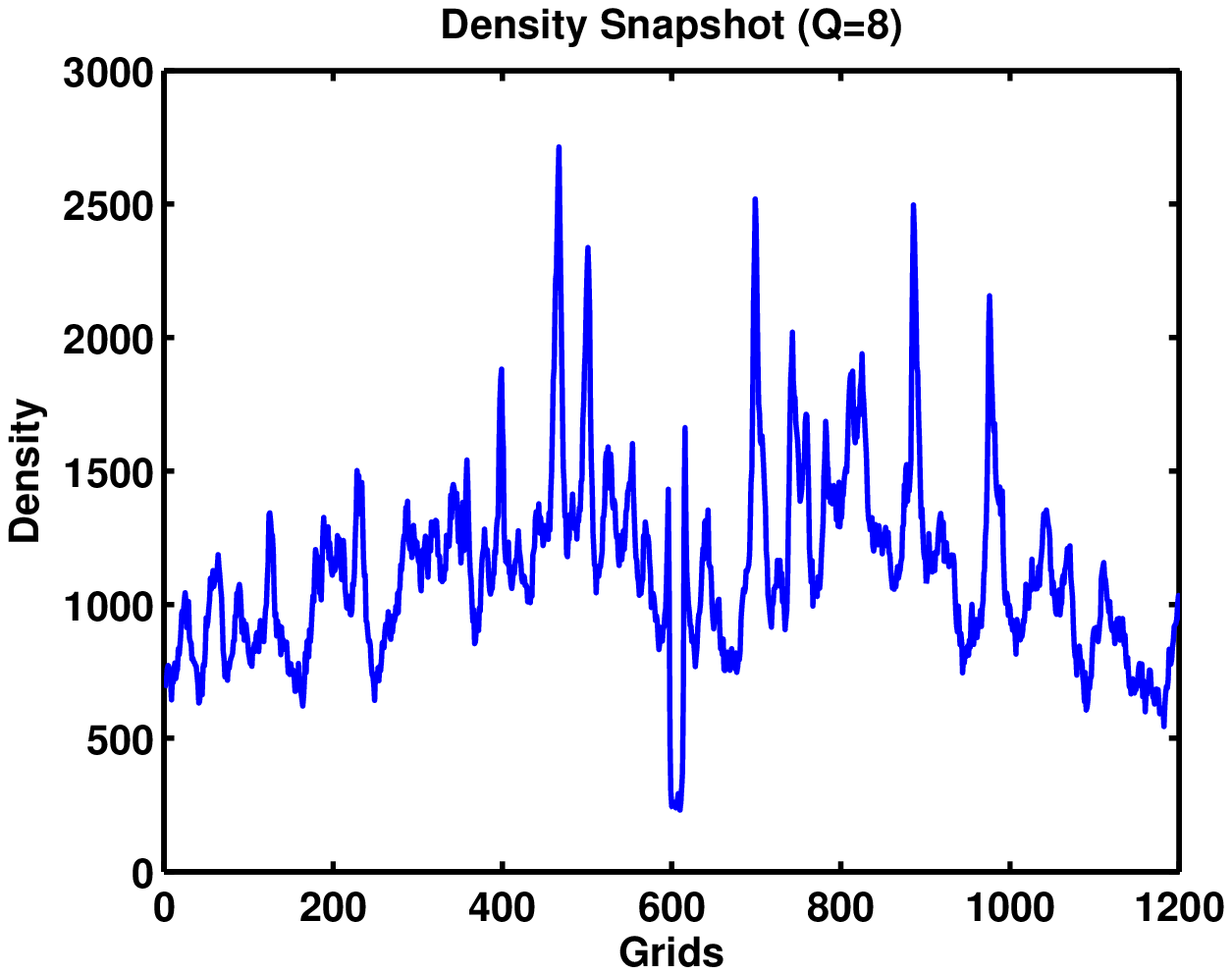}
  \caption{\small The top panel represents snapshots of the bulk flow velocity at the periods of simulation time Q=3, 5, 7.
  The low panel represents snapshots of the flow density at the periods of simulation time Q=4, 6, 8.}
  \label{fig2}
 \end{figure}
The total simulation time $T_{max}$=2400 are divided into 10 periods of time represented by Q=1, 2, 3...10 sequently.
Fig.\ref{fig2} shows a group of bulk flow velocity snapshots in sequence of Q=3,5,7 and  density snapshots in sequence
of Q=2, 4, 8. In the top panel, the bulk flow velocity with initial $u0_{1}$ and $u0_{2}$ in the upstream region
evolute into the downstream with bulk flow speed of zero. From the Q=3, 5, to 7, the downstream region extend from the
both boundaries to the center of the simulation box. Two shock fronts propagate from the both boundaries to the center
of the box with opposite shock velocities $v_{sh1}$ and $v_{sh2}$, symmetrically. In front of the shocks, the velocity
profiles have gradual slopes, which represent the shock precursors caused by the energetic particles back-reaction on
the shock. When the two shock fronts approach  more and more close, the two precursors overlap and have an interaction
between the energetic particles extracted from the both downstream regions. This interaction process may vary the
regular energy spectrum. In the low panel, the bulk flow density with an initial number dennsity $n_{0}$ per grid
evolute into downstream region with a higher density. The downstream region show a higher density with several times of
that in the upstream region. Noted that both downstream regions have little lower densities in time period of Q=8 than
Q=6 and 4, because our simulation model applying an adapter density with the simulation time. For the simply
computation, we apply a temporally density with a reduction of density $dn$=16 in each period of time Q (i.e. adapter
density reduce 1 particle from initial density $n_{0}$ per 15 simulation time units). Similarly, the density profiles
can also produce an interaction on both precursor regions when the two shocks become close enough. This may induce the
precursor energy spectrum with the same spectrum in the downstream region at the end of the simulation.

\begin{figure}[t]\center
    \includegraphics[width=2.0in, angle=0]{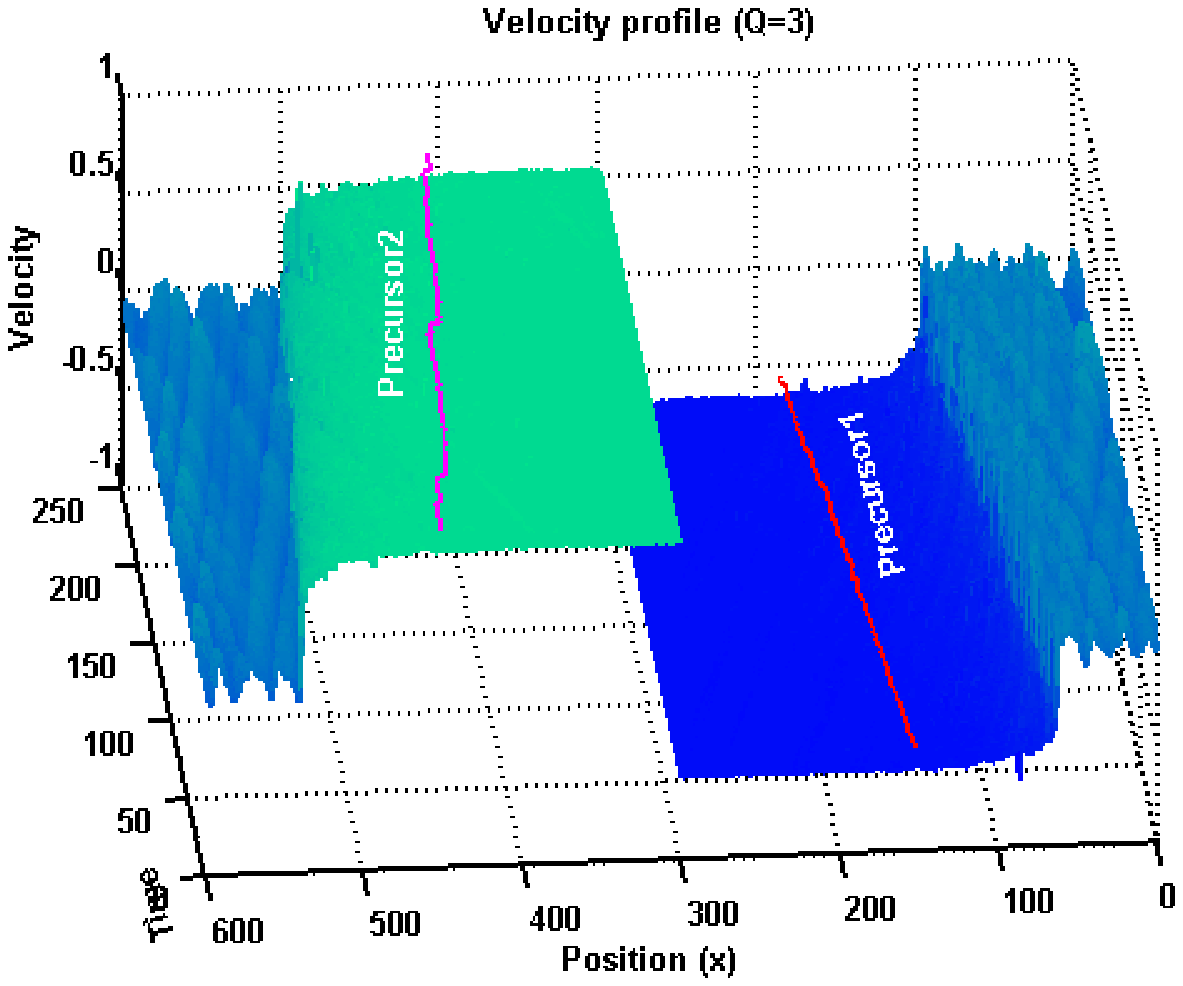}
    \includegraphics[width=2.0in, angle=0]{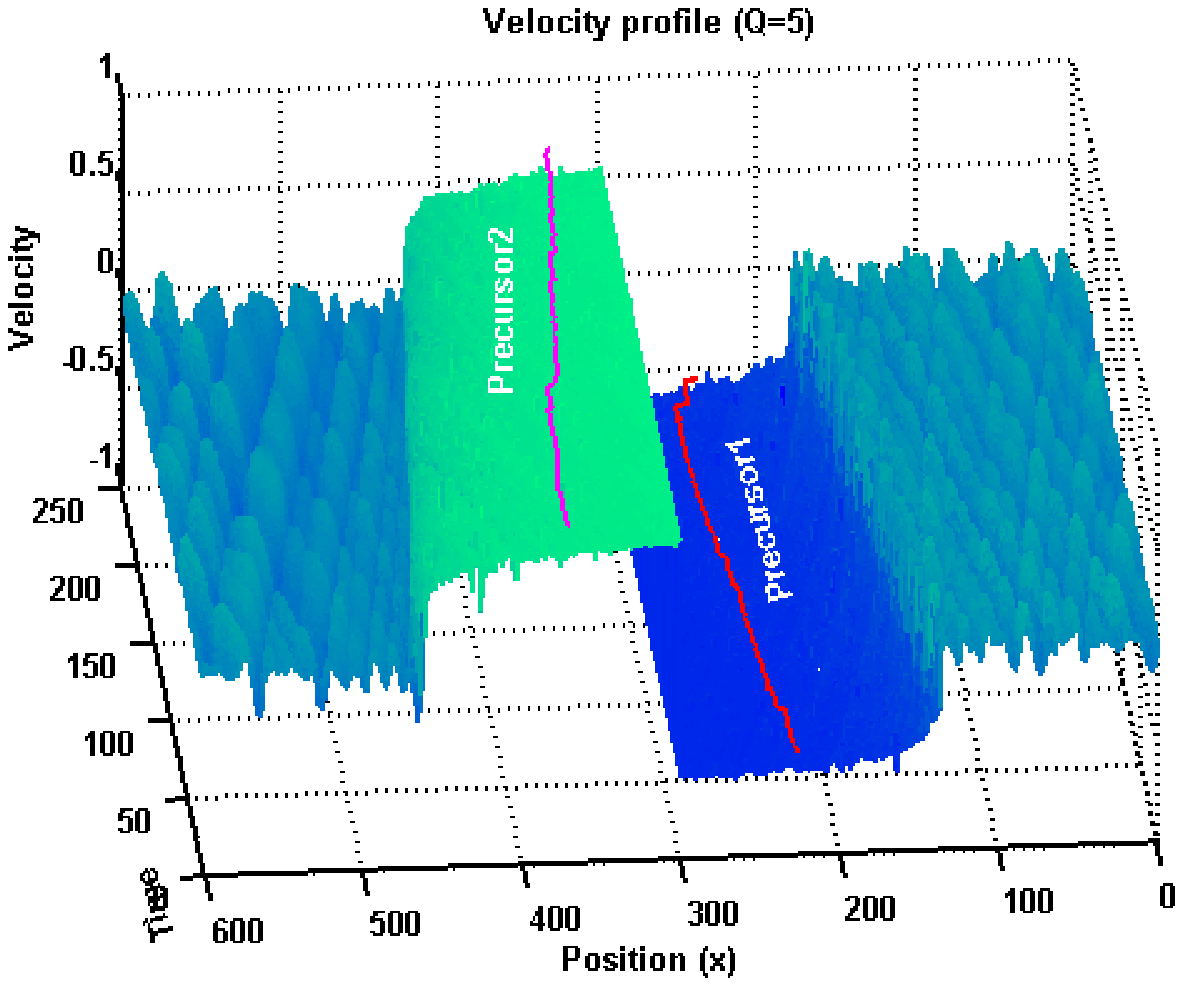}
    \includegraphics[width=2.0in, angle=0]{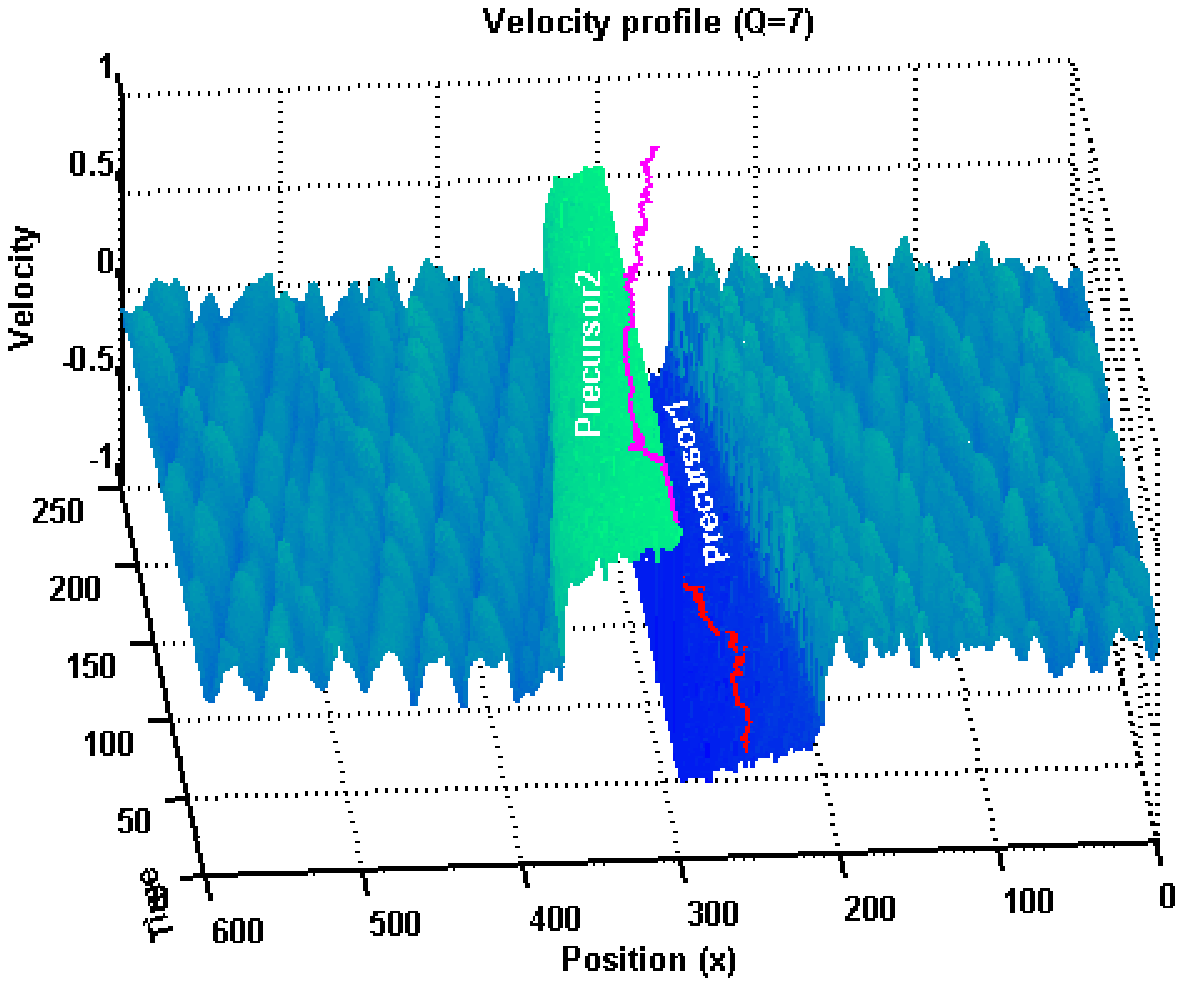}\\
    \includegraphics[width=2.0in, angle=0]{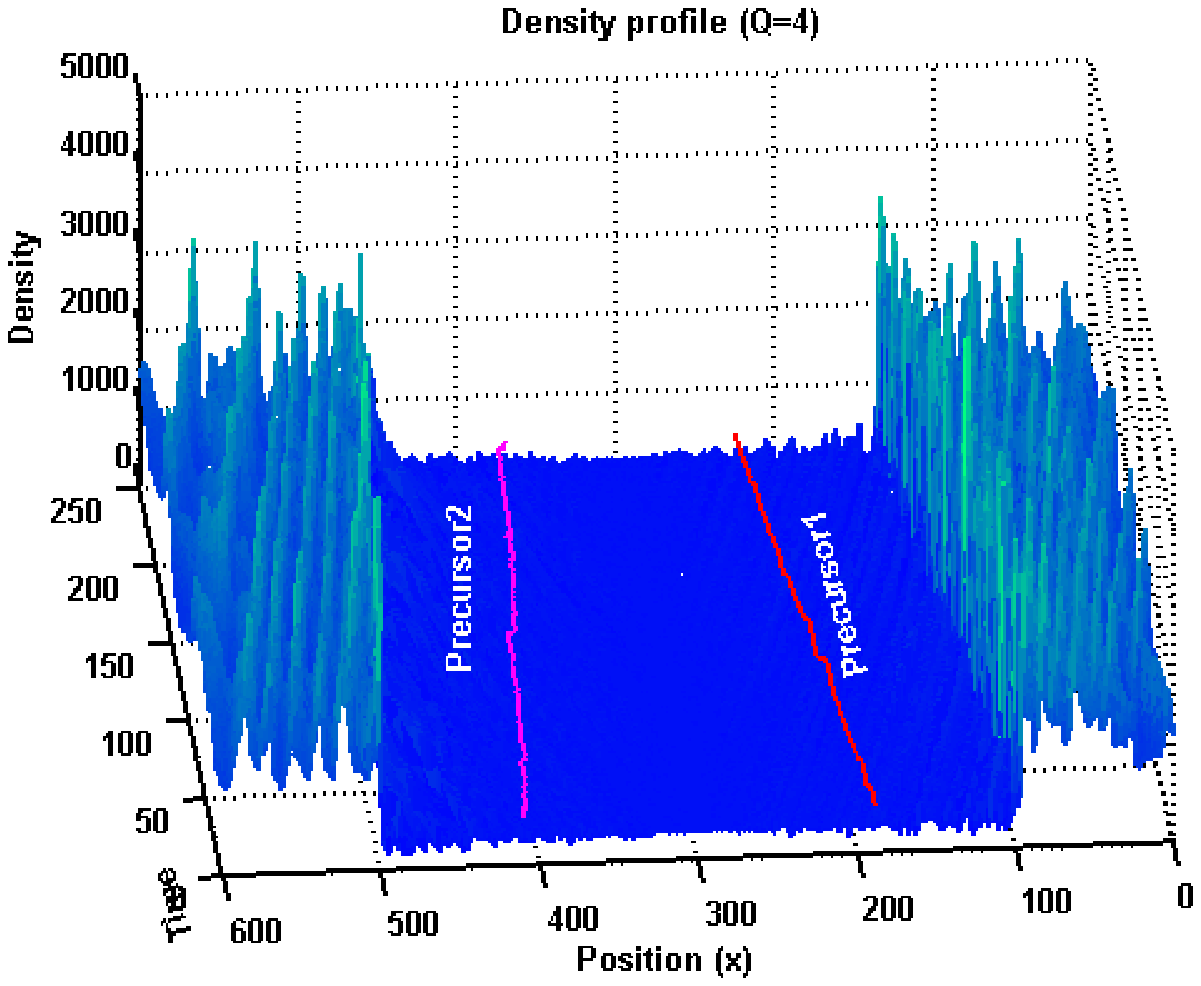}
    \includegraphics[width=2.0in, angle=0]{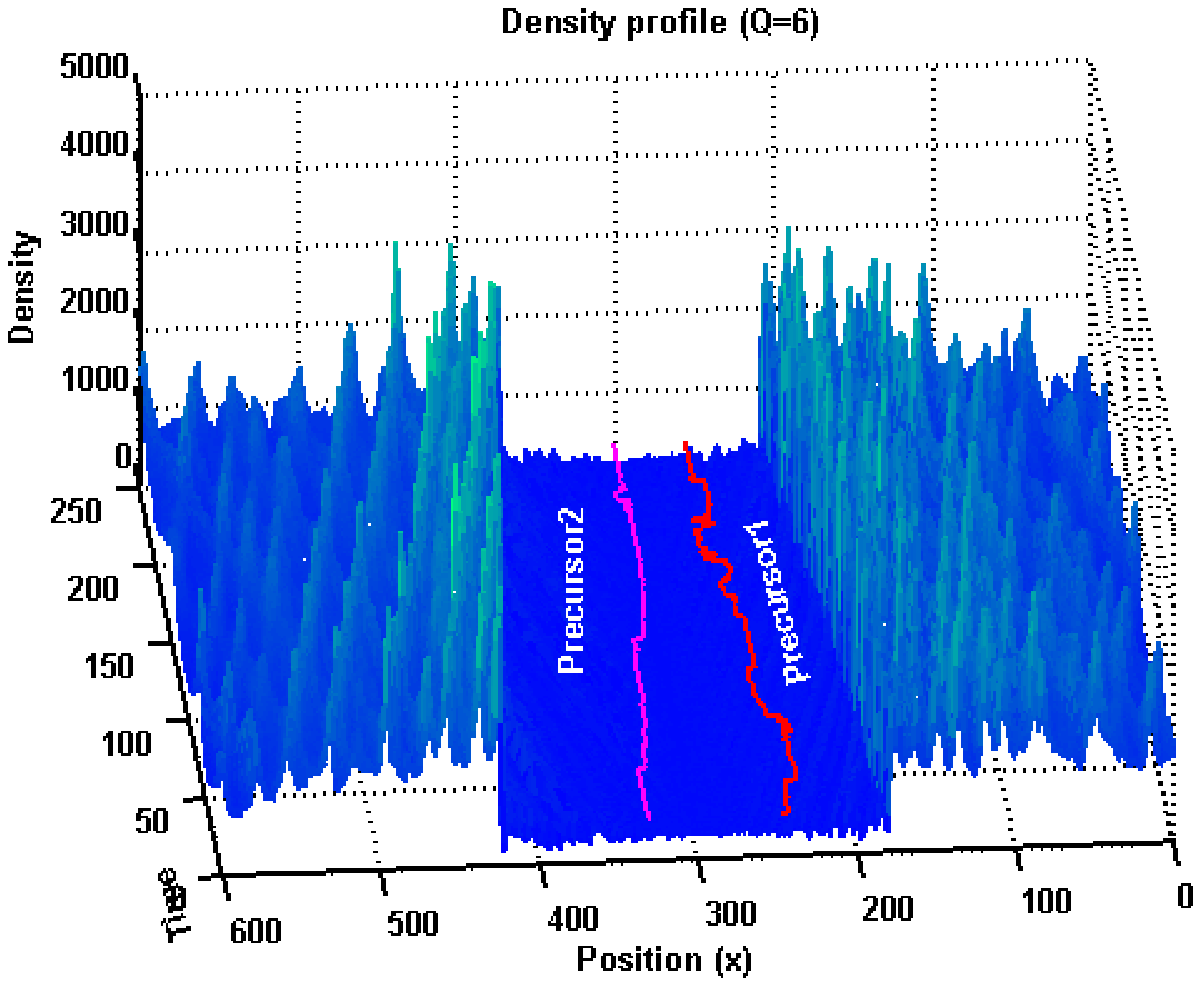}
    \includegraphics[width=2.0in, angle=0]{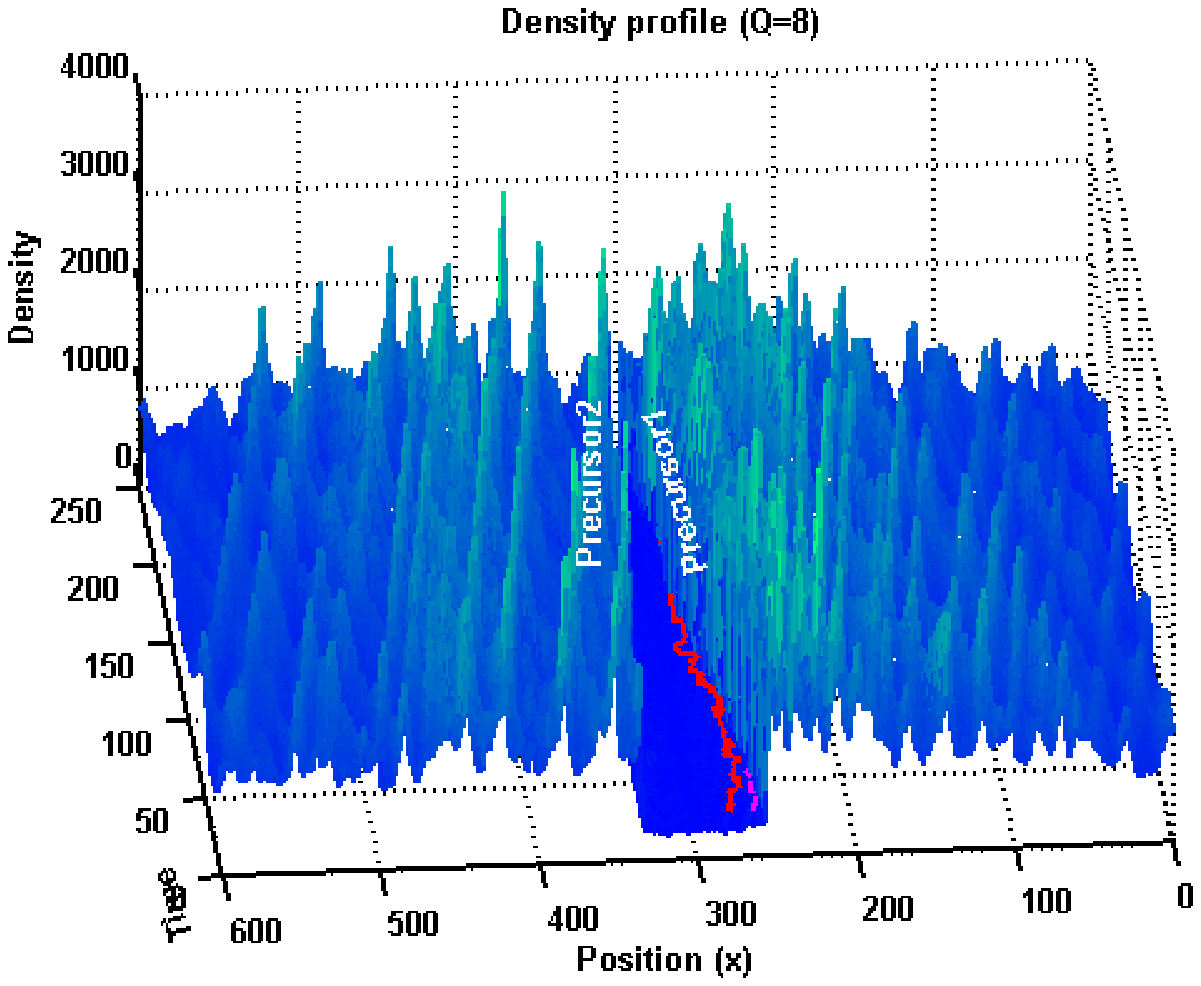}
    \caption{\small The top panel represents profiles of the bulk flow velocity at the periods of simulation time Q=3,5,7.
  The low panel represents profiles of the bulk flow density at the periods of simulation time Q=4,6,8.}
  \label{fig3}
\end{figure}

Fig.\ref{fig3} shows a group of the bulk flow velocity profiles at the periods of Q=3, 5, 7 and density profiles at the
periods of Q=4, 6, 8. Top panel shows a series of the bulk flow velocity profiles in the position with the time. In the
laboratory reference frame, the bulk flow velocity in both of the downstream regions are equal to zero, the bulk flow
velocities in two upstream regions show an opposite movement in the middle of simulation box. The solid lines in both
upstream regions represent the precursor position, respectively. The top right mesh plot shows the two precursor
regions overlapping with time. This overlapping processes play two roles on the energetic particles: (i) shock 2 with a
positive precursor bulk velocity decelerate the energetic particles jumped from shock 1 with a negative precursor bulk
velocity. (ii) In the contrary, shock 1 with a negative precursor bulk flow velocity re-accelerate the energetic
particles dropped from shock 2 with a positive precursor bulk flow velocity. Lower panel shows a series of density
profiles in the position with the time. The three mesh plots indicate the higher density in the downstream regions
shorten the upstream regions with a lower density from the periods of Q=4, 6, to 8. The right plot shows two precursor
regions are mixed together with a hybrid density.

\subsection{Particle Acceleration}

\begin{figure}[t]\center
    \includegraphics[width=2.3in, angle=0]{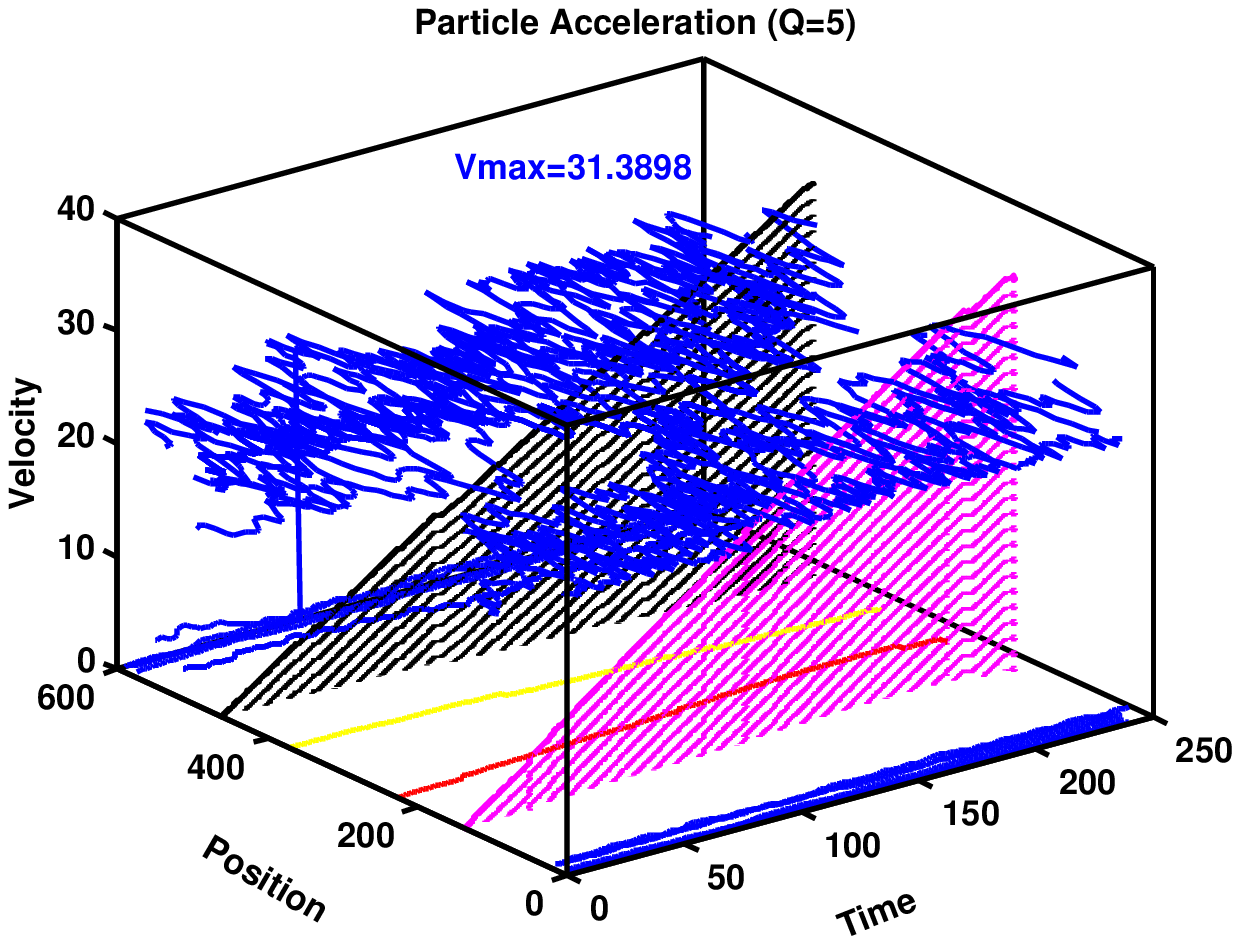}
    \includegraphics[width=2.3in, angle=0]{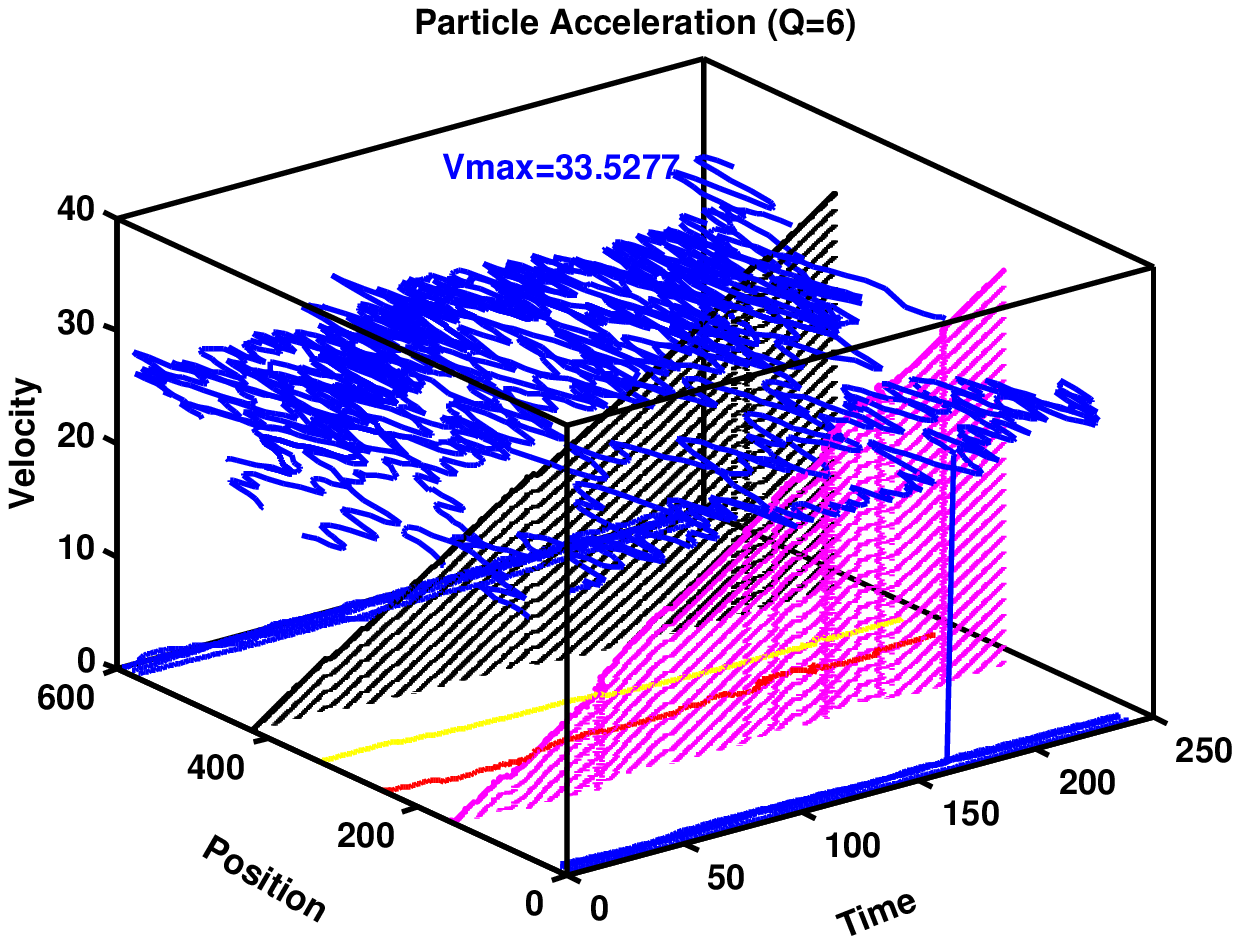}\\
    \includegraphics[width=2.3in, angle=0]{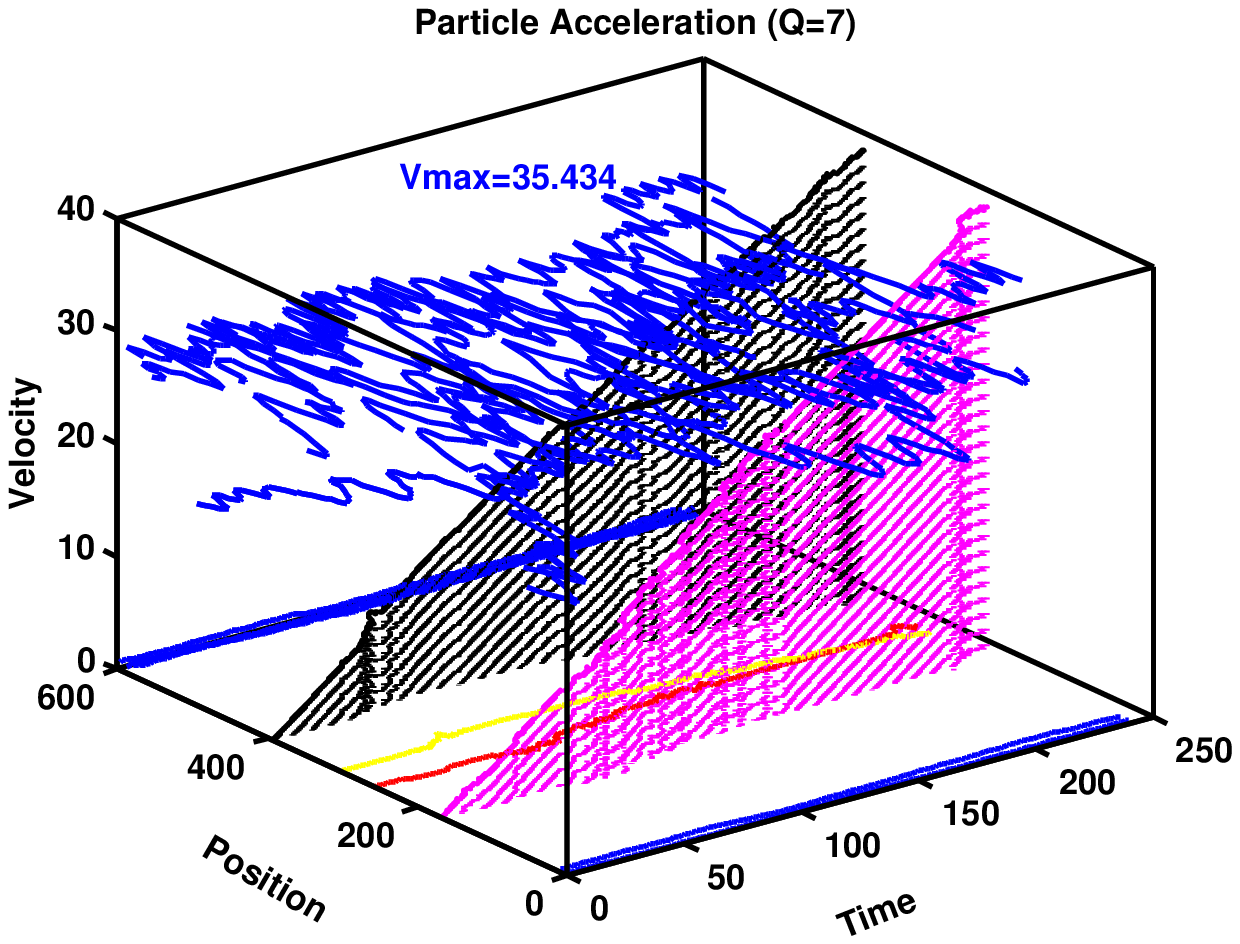}
    \includegraphics[width=2.3in, angle=0]{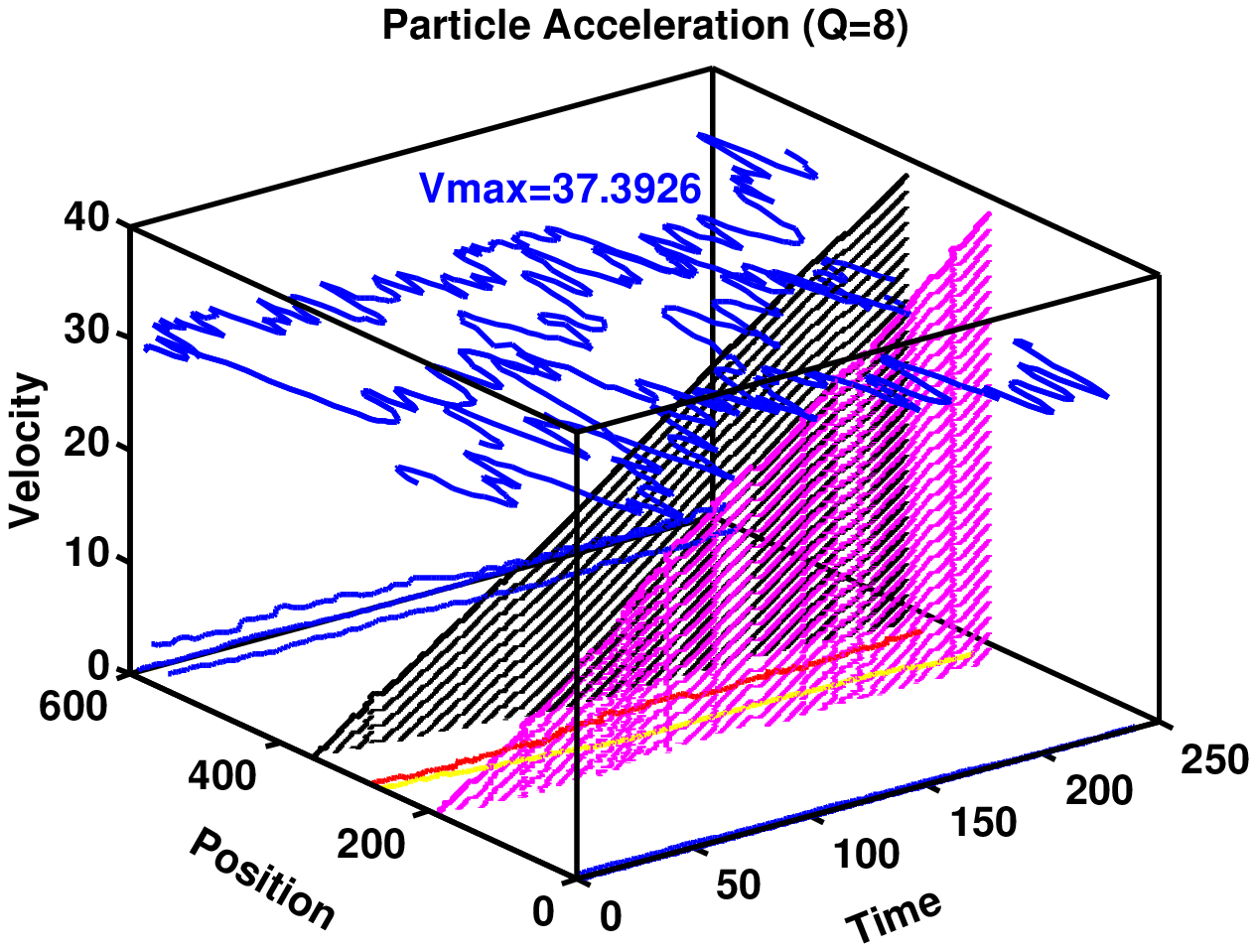}\\
    \includegraphics[width=2.3in, angle=0]{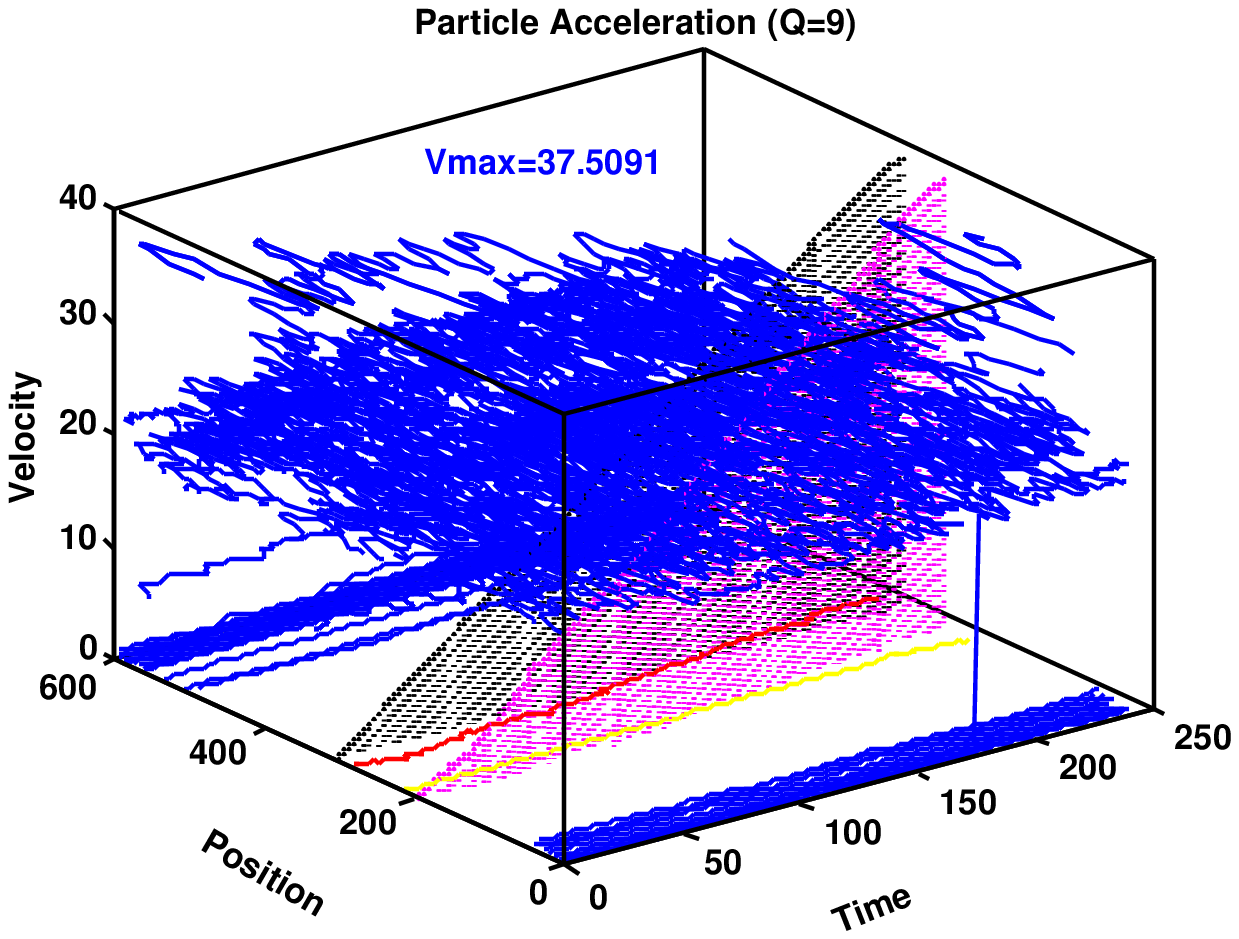}
    \includegraphics[width=2.3in, angle=0]{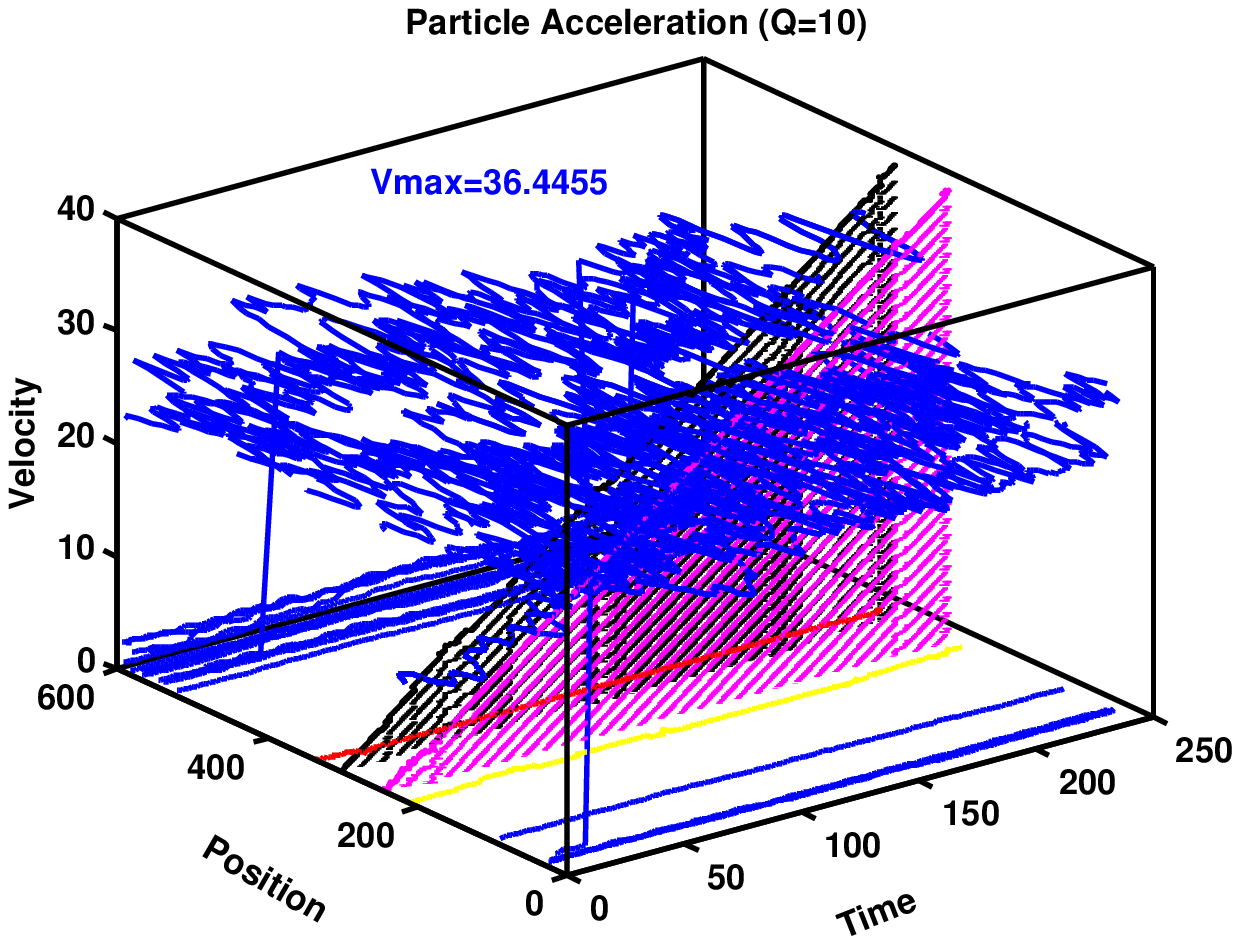}
    \caption{\tiny This group of plots represent particles acceleration in double shocks at the periods of simulation
    time Q=5, 6, 7, 8, 9, and 10. These curves in each plot represent particle trajectories. These trajectories show
    local velocity evolutions of those particles with position and time.}
  \label{fig4}
\end{figure}

Fig.\ref{fig4} shows a group of particle acceleration processes in periods of Q=5, 6, 7, 8, 9, and 10. There are a part
of particles extracted from the total simulation box in corresponding plot. In each pot, two triangle shadow areas
represent the two shock fronts. Some of particles rotating in each shock front trace their trajectories with energy
gains from the lower velocity to the higher velocity due to multiple scattering cycles on each shock front. Another
particles in the downstream regions indicate they have no energy gains without velocity increasing. Maximum particle
velocities $V_{max}$ denoted with values for 31.3898, 33.5277, 35.4340, 37.3926, 37.5091, and 36.4455 are calculated in
periods of time Q=5, 6, 7, 8, 9, and 10, respectively. These maximum velocities show that these particles keep
accelerating with the time to achieve an saturation at a certain time, then they exhibit deceleration processes at the
end of the simulation. In the periods of time between Q=8 and Q=9, two opposite shock fronts approach more close
enough, it make the energetic particles produced by one shock cross into another shock each other. Simultaneously,
these two precursor regions can make the energetic particles mixing in a hybrid precursor region. Because there exist
two different modifications of the precursor bulk flow velocities, the energetic particles produced by one shock region
may induce either deceleration or re-acceleration in another shock region. These mixture may break a smooth single
energy spectral power law followed by a single shock model.


\subsection{Energy Spectra}

 \begin{figure}[t]
\center
        \includegraphics[width=2.0in,angle=0]{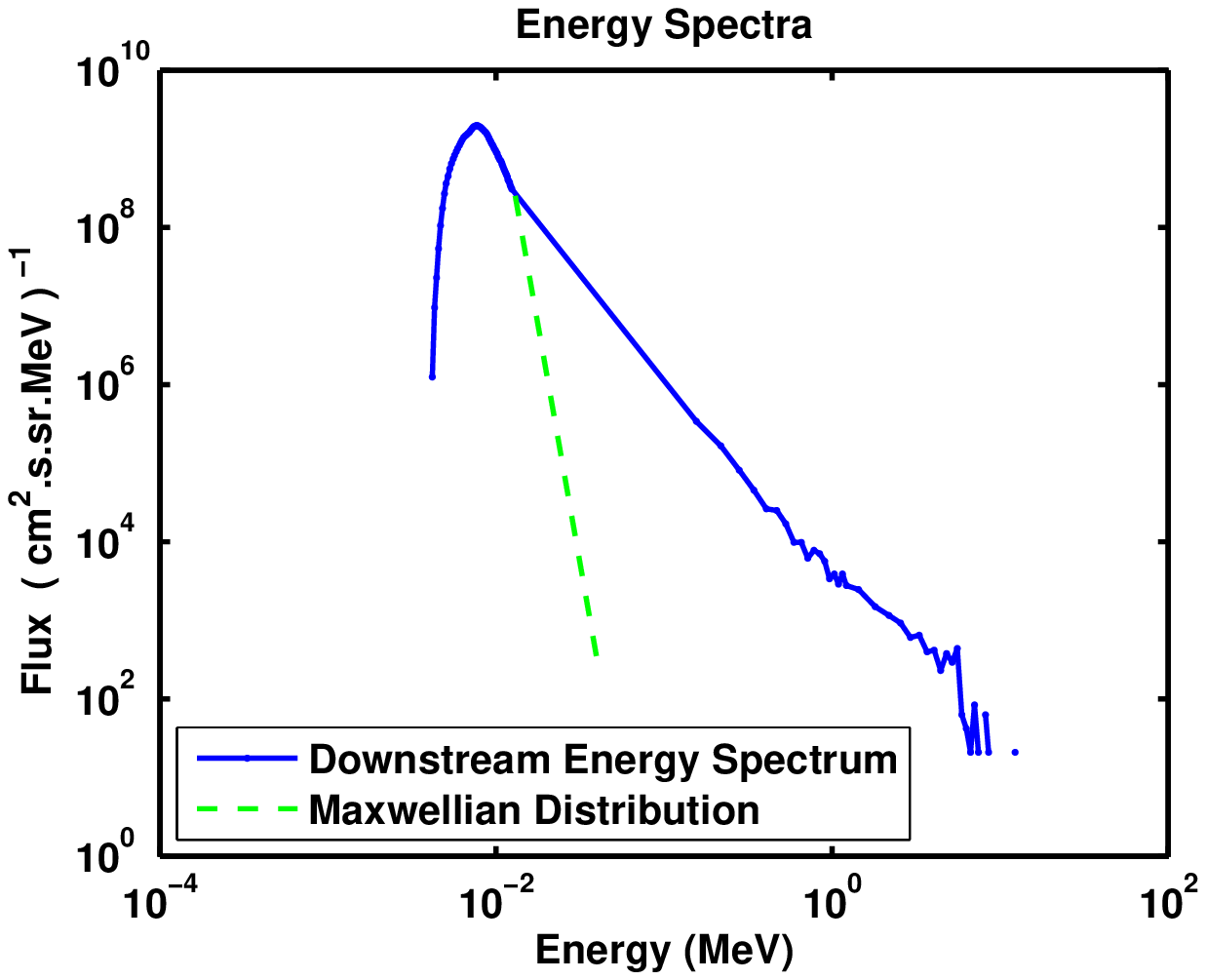}
        \includegraphics[width=2.0in,angle=0]{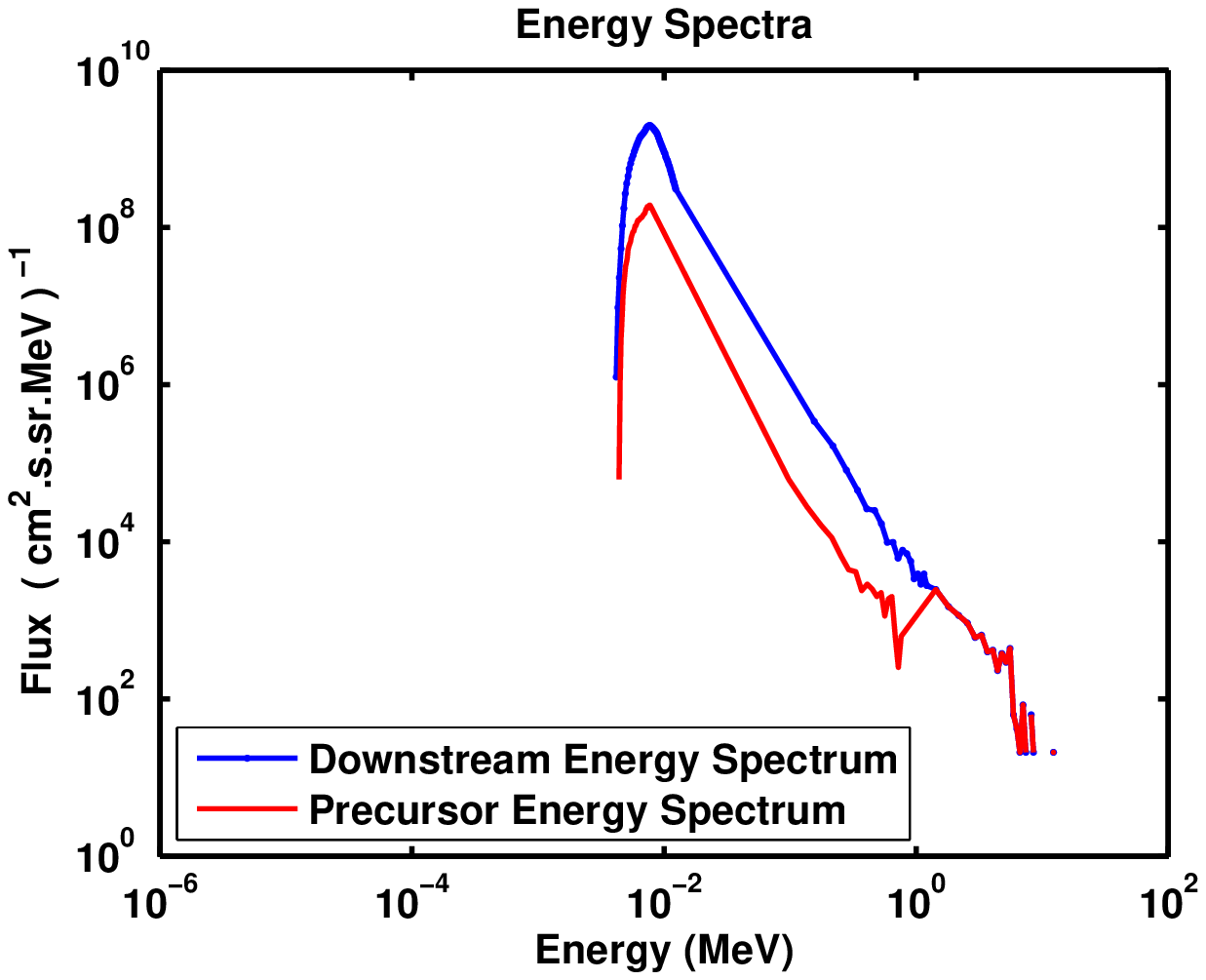}\\
        \includegraphics[width=2.0in,angle=0]{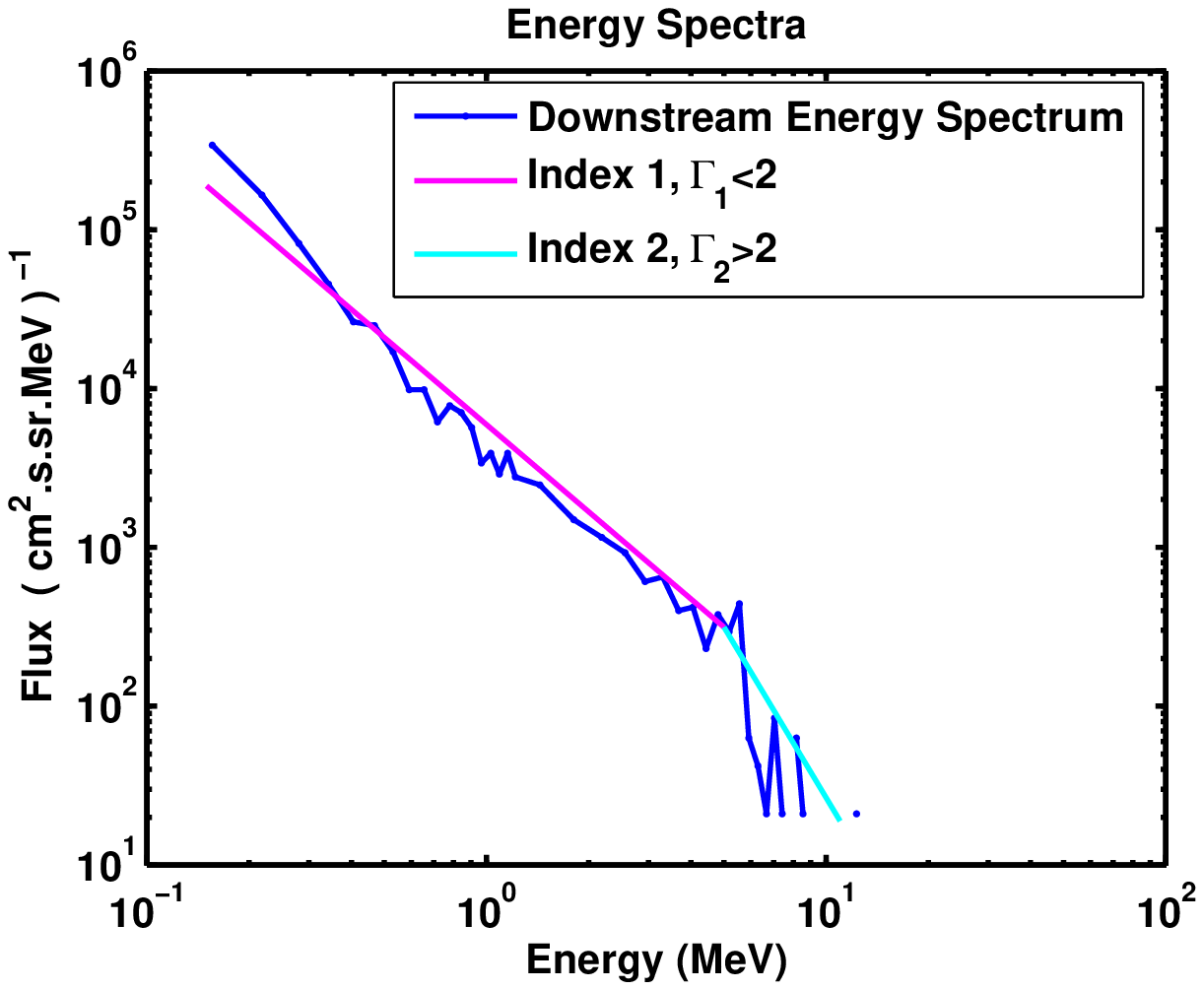}
        \includegraphics[width=2.0in,angle=0]{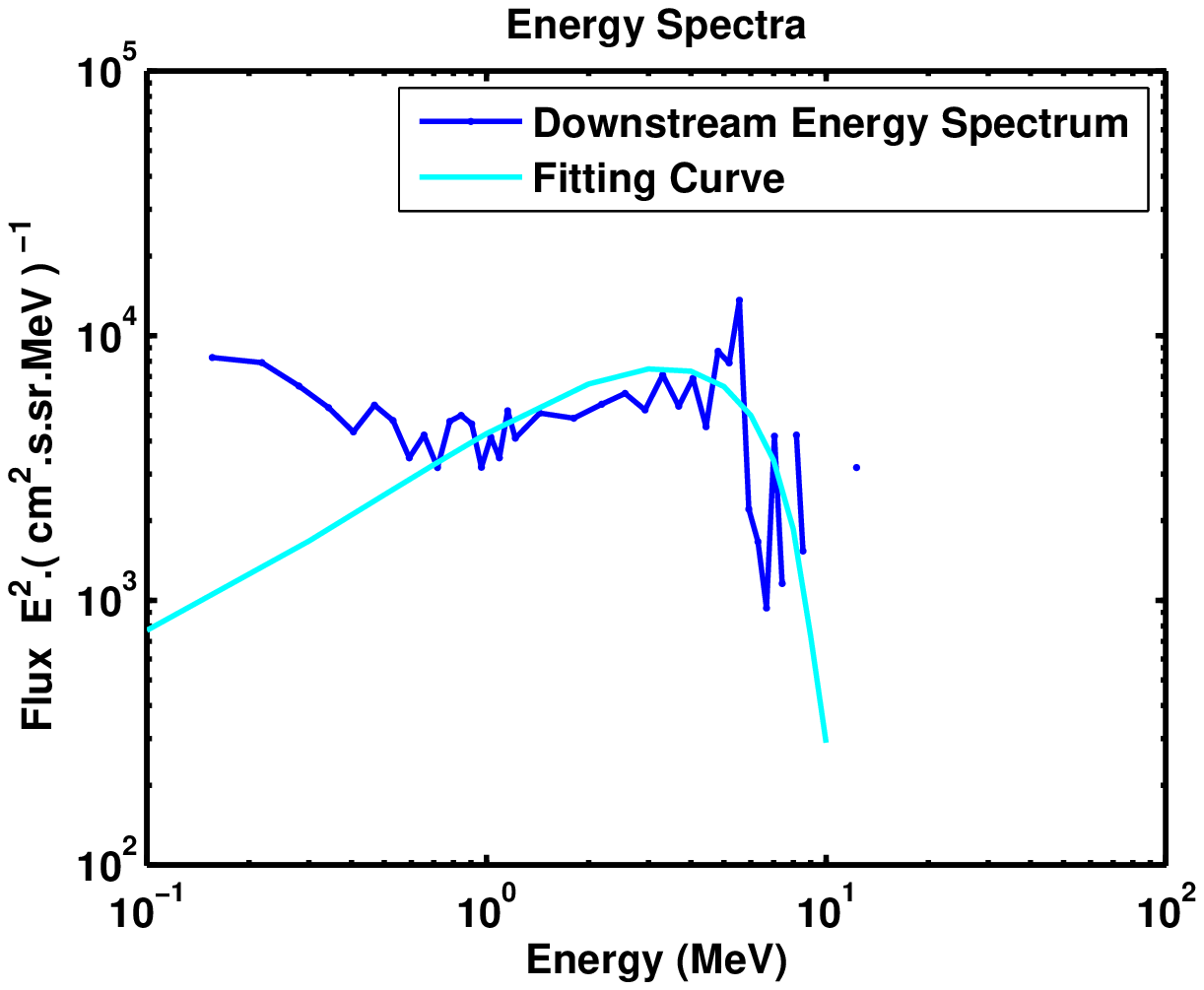}\\
        \includegraphics[width=2.1in,angle=0]{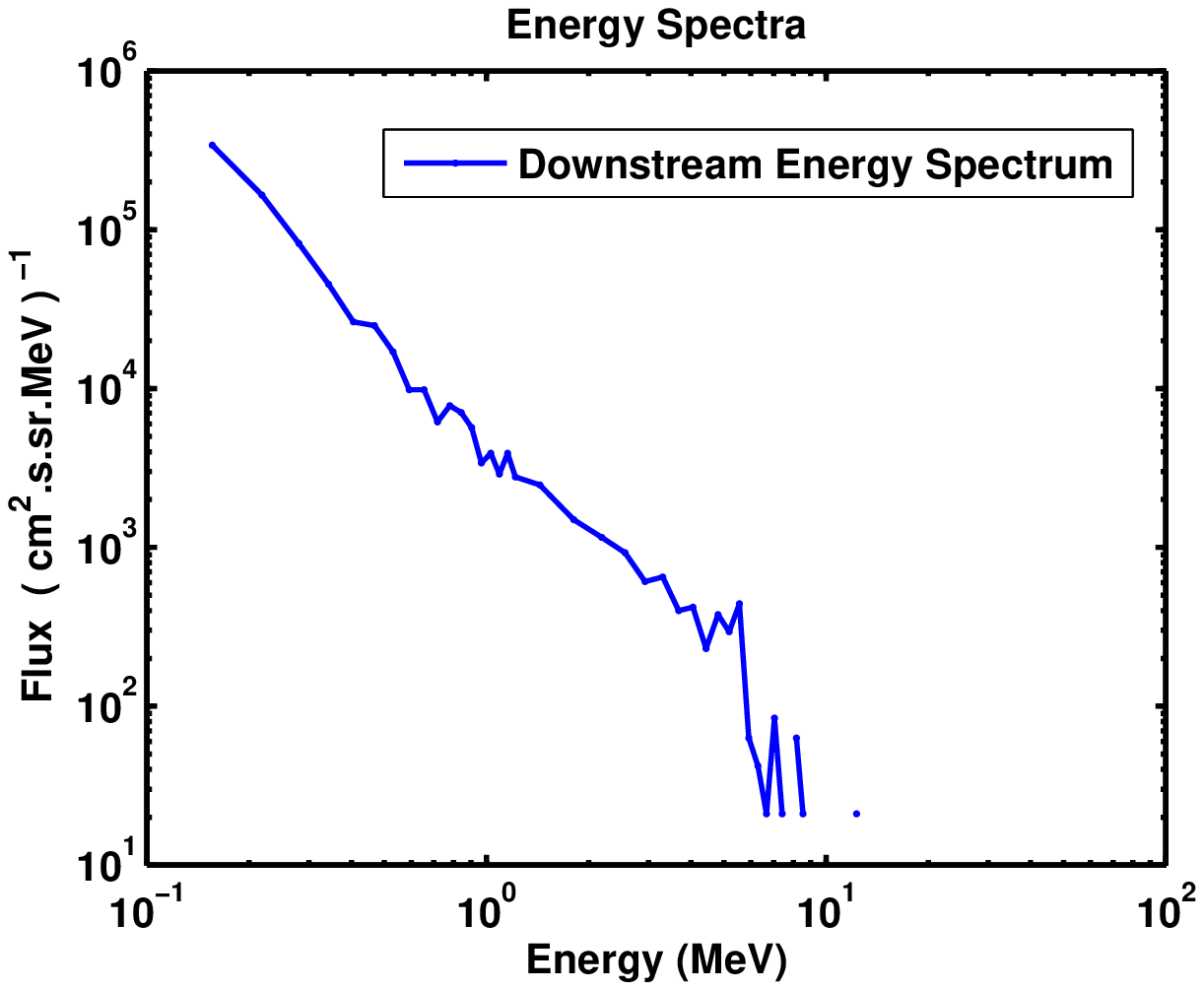}
        \includegraphics[width=1.7in,angle=0]{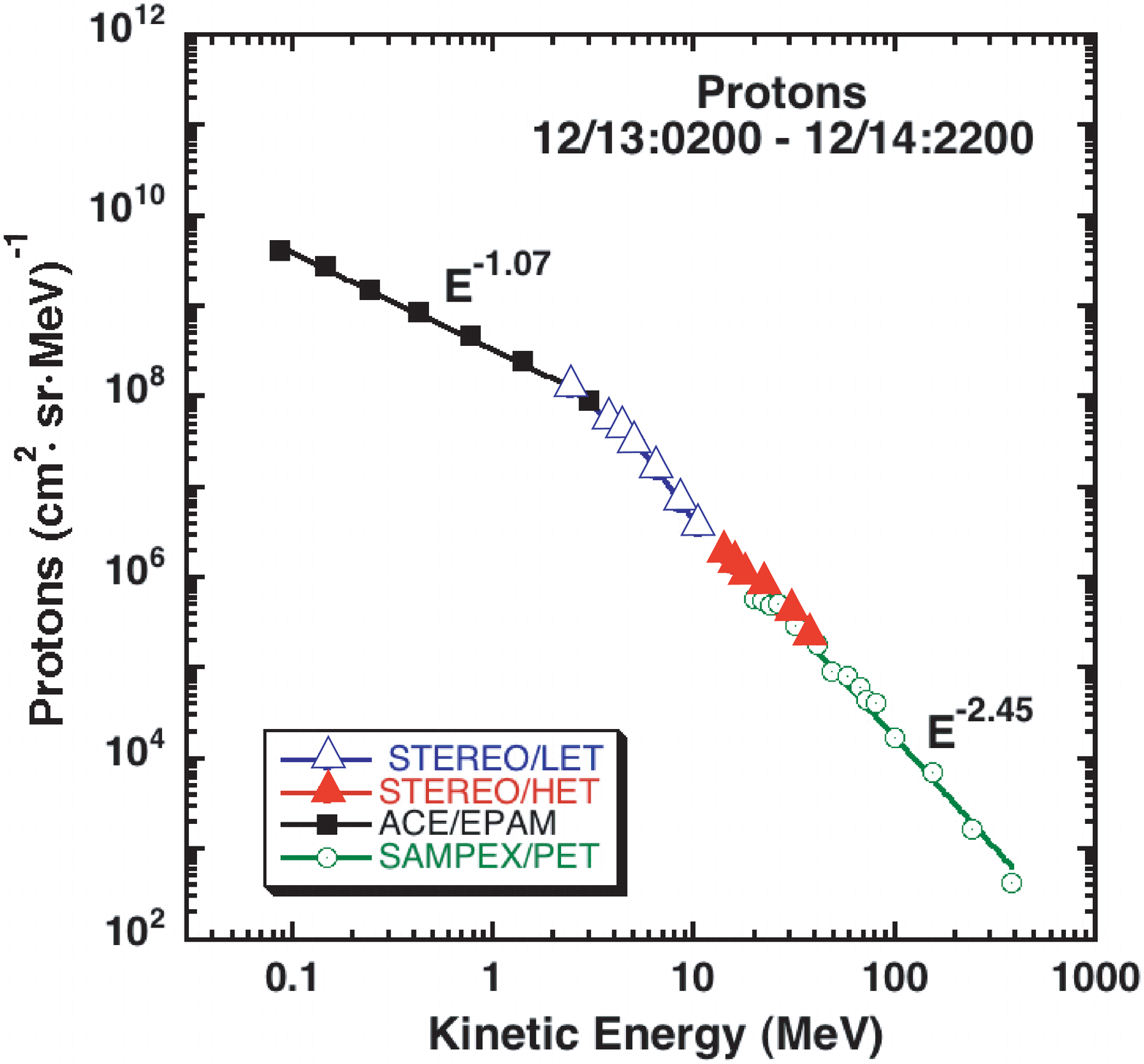}
  \caption{\tiny In top left plot, the solid line represents the proton spectrum with a ``break" power law  obtained from
  the downstream region at the end of simulation, the dashed line represents the thermal distribution in downstream
  region.  In top right plot,  the red line shows energy spectrum in the precursor region;
  the blue line shows energy spectrum in the downstream region. The middle left plot is the higher energy part of the
  top left plot. The pink line and cyan line show an energy spectral ``break" at $\sim$5MeV. The middle right plot
  also indicates double energy spectral indices. Cyan curve fitted the simulation data shows one index in lower energy
  range is less than 2, and another index in higher energy range is more than 2. The low left plot shows the energy
  spectra in downstream region purely with comparison of observed spectrum from the multiple spacecraft. The flux in left
  plot is statistic in per second, it is consist with the flux of the right plot in period of time 44 hours.}
  \label{fig5}
 \end{figure}


Fig.\ref{fig5} shows a group of proton spectra calculated from simulation box at the end of the simulation. Each plot
shows an energy  spectral ``break" at $\sim$5MeV. The blue curve in top left plot represents the energy spectrum with a
Maxiwellian peak at a few keV and with an extended power law energy spectral``tail" in downstream region. The energy
spectral ``break" occurs at the energy of $\sim$ 5MeV.  The dashed line indicates the potential Maxwellian distribution
in the shocked downstream. The top right plot represents two energy spectra in upstream and downstream regions,
respectively. The pink color curve denotes the energy spectrum in upstream region and the blue color curve denotes the
energy spectrum in downstream region. From the energy range of $\sim$ 1MeV to  $\sim$ 10MeV, the energy spectrum in
upstream region with the same shape of the energy spectrum in downstream. Both of them appear energy spectral ``breaks"
at $\sim$ 5MeV. For convenience to study the energy spectral ``break", we extract a part of energy spectrum from the
total downstream energy spectrum as shown in the middle left plot. The pink line represents the harder energy spectrum
with an index less than 2, and the cyan line represents the softer energy spectrum with an index more than 2. The
``break" indicate that there exist a double power law spectrum. The middle right plot gives an energy spectral shape in
the flux of $E^{2}\cdot F(E)$ representation, in which the lower energy spectrum exhibits a descent slope shape, and
the higher energy spectrum exhibits a dropped slope shape. The fitting curve clearly shows that there exist a double
power-law with a ``break" at $\sim$ 5MeV. The low panel presents a comparison of the simulated energy spectrum with the
observed energy spectrum. The low left plot is the simulated energy spectrum in the downstream region purely; the low
right plot is the observed energy spectrum in the downstream region too \citep{mewaldt08}. The difference of the flux
between two energy spectra is caused by different integrating time. The simulated flux is integrated in one second, but
the observed flux is the sum of the period from 2006 Dec 13, 02:00 to Dec 14, 22:00, which is equivalent to
1.584$\times 10^{5}$s. So the simulated and the observed fluxes are well agreement in the normalized time.

\section{Summaries and Conclusions}\label{sec-summary}
In summary, we do simulation of the  converged two shocks for obtaining the proton spectrum directly. Comparing with
the observed energy spectrum from multiple spacecraft. Simulated energy spectrum exhibits the consist ``break" at the
energy $\sim$5MeV. Although the numerical computation is very expensive, we obtain the highest energy spectral ``tail"
up to $\sim$10MeV. We have updated the results of the maximum particle energy, which can not predict the energy
spectral ``break" in a single shock model. Why the previous efforts can not predict the energy spectral ``break" ?
There would be two reasons: (i) the single shock model can not provide enough energy to transfer the energetic
particles to the highest energy up to $\sim $10MeV. (ii) The shortage of the interaction mechanism make it impossible
to re-accelerate or decelerate the energetic particles. However, the converged shocks model can satisfy these two
essential conditions. Firstly, the converged shocks model can provide enough energy injection than a single shock model
to produce the highest energy spectral ``tail". Secondly, it can also provide an interaction mechanism to break the
standard single power law energy spectrum formed by a single shock model. In converged two shocks, those energetic
particles produced in each individual shock can mix together into the hybrid precursor region when two shocks approach
more and more close. It is just these opposite precursor bulk flow velocities present the variation of the energetic
particles distribution and produce an energy spectral``break"  at the higher energy tail.


%

\acknowledgements{Present work is supported by Xinjiang Natural Science Foundation No. 2014211A069. This work is also
funded by CAS grant KLSA201511, the Key Laboratory of Modern Astronomy and Astrophysics (Nanjing University), Ministry
of Education, and China Scholarship Council (CSC). Simultaneously, authors thank the support from Supercomputer Center
of University of Arizona. In addition, authors are also appreciate Profs. Hongbo Hu and Hong Lu in Institute of High
Energy Physics of Chinese Academy of Sciences for their many helpful discussions.}

\end{document}